\documentclass[11pt]{article}

\title{Two-step estimation of a multivariate L\'evy process}

\author{Habib Esmaeili
\thanks{Center for Mathematical Sciences, Technische Universit\"at M\"unchen,  Boltzmannstra\ss e 3, 85748 Garching, Germany,
email: esmaeili@ma.tum.de, http://www-m4.ma.tum.de/} \and Claudia
Kl\"uppelberg\footnotemark[0]
\thanks{Center for Mathematical Sciences, and Institute for Advanced Study, Technische Universit\"at M\"unchen, Boltzmannstra\ss e 3, 85748 Garching, Germany,
email: cklu@ma.tum.de, http://www-m4.ma.tum.de/}}

\usepackage{epsfig,psfrag}
\usepackage{graphicx}
\usepackage{amssymb}
\usepackage{amsfonts}
\usepackage{color}
\usepackage{amsmath}
\usepackage{amsthm}
\usepackage{multirow}
\usepackage{rotating}

\numberwithin{equation}{section}

\newtheorem{theorem}{Theorem}[section]
\newtheorem{lemma}[theorem]{Lemma}
\newtheorem{remark}[theorem]{Remark}
\newtheorem{example}[theorem]{Example}
\newtheorem{proposition}[theorem]{Proposition}
\newtheorem{definition}[theorem]{Definition}
\newtheorem{corollary}[theorem]{Corollary}
\newtheorem{fig}[theorem]{Figure}

\newcommand{\bthe}{\begin{theorem}}
\newcommand{\ethe}{\end{theorem}}

\newcommand{\ben}{\begin{enumerate}}
\newcommand{\een}{\end{enumerate}}

\newcommand{\beq}{\begin{equation}}
\newcommand{\eeq}{\end{equation}}

\newcommand{\ble}{\begin{lemma}}
\newcommand{\ele}{\end{lemma}}

\newcommand{\bde}{\begin{definition}}
\newcommand{\ede}{\end{definition}}

\newcommand{\bco}{\begin{corollary}}
\newcommand{\eco}{\end{corollary}}

\newcommand{\bpr}{\begin{proposition}}
\newcommand{\epr}{\end{proposition}}

\newcommand{\brem}{\begin{remark}\rm}
\newcommand{\erem}{\end{remark}\rm}

\newcommand{\bexam}{\begin{example}\rm}
\newcommand{\eexam}{\end{example}}

\newcommand{\bfi}{\begin{fig}}
\newcommand{\efi}{\end{fig}}

\newcommand{\btab}{\begin{tab}}
\newcommand{\etab}{\end{tab}}

\newcommand{\beqq}{\begin{equation}}
\newcommand{\eeqq}{\end{equation}}

\newcommand{\beao}{\begin{eqnarray*}}
\newcommand{\eeao}{\end{eqnarray*}\noindent}

\newcommand{\beam}{\noindent\begin{eqnarray}}
\newcommand{\eeam}{\end{eqnarray}\noindent}

\newcommand{\barr}{\begin{array}}
\newcommand{\earr}{\end{array}}

\newcommand{\bdis}{\begin{displaymath}}
\newcommand{\edis}{\end{displaymath}\noindent}

\newcommand{\eproof}{\halmos}

\def\bbr{{\Bbb R}}
\def\bbn{{\Bbb N}}

\def\bbp{{\Bbb P}}
\def\bbe{{\Bbb E}}

\def\calf{{\mathcal{F}}}

\newcommand{\Cov}{{\rm Cov }}

\newcommand{\stp}{\stackrel{P}{\rightarrow}}
\newcommand{\std}{\stackrel{d}{\rightarrow}}
\newcommand{\stas}{\stackrel{\rm a.s.}{\rightarrow}}

\newcommand{\nto}{n\to\infty}

\newcommand{\al}{{\alpha}}

\newcommand{\de}{{\delta}}
\newcommand{\thet}{{\theta}}

\newcommand{\la}{{\lambda}}

\newcommand{\ga}{{\gamma}}

\newcommand{\eps}{\varepsilon}
\newcommand{\var}{{\rm Var}}
\newcommand{\cov}{{\rm Cov}}
\newcommand{\corr}{{\rm Corr}}

\newcommand{\leps}{\lambda^{(\eps)}}
\newcommand{\lepss}{\lambda_{**}^{(\eps)}}
\newcommand{\lepsd}{\lambda^{(\eps)\|}}

\newcommand{\ovPieps}{\ov{\Pi}^{(\eps)}}

\newcommand{\lx}{\lambda^{(\xi)}}

\newcommand{\ov}{\overline}
\newcommand{\wh}{\widehat}
\newcommand{\wt}{\widetilde}
\newcommand{\lp}{L\'evy process}
\newcommand{\lm}{L\'evy measure}
\newcommand{\lc}{L\'evy copula}

\newcommand{\bm}{\boldmath}

\newcommand{\bS}{\mathbf S}
\newcommand{\bzero}{\mathbf 0}

\newcommand{\bx}{\mathbf x}

\newcommand{\by}{\mathbf y}

\newcommand{\bJ}{\mathbf J}

\newcommand{\fC}{\mathfrak C}
\newcommand{{\bth}}{\mbox{\bm {$\theta$}}}
\newcommand{{\btheta}}{\mbox{\bm {$\theta$}}}

\newcommand{\p}{\partial}

\newcommand{\argmax}{{\rm argmax}}

\newcommand{\halmos}{\quad\hfill\mbox{$\Box$}\\}  

\newcommand{\h}[1]{{\color{red}#1}}

\begin{document}


\maketitle

\begin{abstract}
Based on the concept of a L\'evy copula to describe the dependence structure of a multivariate L\'evy process we present a new estimation procedure. We consider a parametric model for the marginal L\'evy processes as well as for the L\'evy copula and estimate the parameters by a two-step procedure. We first estimate the parameters of the marginal processes, and then estimate in a second step only the dependence structure parameter. For infinite L\'evy measures we truncate the small jumps and base our statistical analysis on the large jumps of the model. Prominent example will be a bivariate stable \lp, which allows for analytic calculations and, hence, for a comparison of different methods. We prove asymptotic normality of the parameter estimates from the two-step procedure and, in particular, we derive the Godambe information matrix, whose inverse is the covariance matrix of the normal limit law. A simulation study investigates the loss of efficiency because of the two-step procedure and the truncation.
\end{abstract}

\vfill

\noindent
\begin{tabbing}
\end{tabbing}

\vspace{1cm}


\vspace{0.5cm}

\newpage

\section{Introduction}\label{s1}

In Esmaeili and Kl\"uppelberg~\cite{EsK2} we presented the maximum
likelihood estimation (MLE) for a bivariate stable subordinator.
We assumed for the marginal subordinators to be both stable with
the same parameters and modeled the dependence structure by a
Clayton L\'evy copula. Estimation was based on observed
jumps larger than some predefined $\eps>0$ in both components
within a fixed interval $[0,t]$. For this model we computed the
 MLEs numerically and proved asymptotic normality for
$\eps\to0$ and/or for $t\to\infty$, respectively.
It is certainly useful to know
that such a procedure works; but for more general models as, for
instance, for higher dimensional models with different marginal
L\'evy processes, this estimation method becomes
computationally very expensive.

Consequently, we present in this paper an alternative, which is a
L\'evy equivalent of the so-called IFM (inference functions for
margins) method, a standard method in multivariate statistics; cf.
Godambe~\cite{God91}, Joe~\cite{Joe:MultiModelsDepConcepts},
Ch.~10, and Xu~\cite{Xu96}, Ch.~2.
The observation scheme as chosen in Esmaeili and
Kl\"uppelberg~\cite{EsK2} was simple in the sense that we only
considered observations with jumps in both components larger than
some $\eps>0$. For this observation scheme, however, the marginally truncated processes are not independent of the L\'evy copula parameter.

The appropriate observation scheme to separate marginal and dependence parameters of the small
 jumps truncated processes requires to consider each component
process separately and observe jumps larger than $\eps$ in each single component.
This results again in a compound Poisson process (CPP), where jumps larger than $\eps$ in both components
are seen as joint jumps, and those jumps with sizes larger than $\eps$ only in one component
(and smaller in the other) are treated as positive jumps in one component and jump size 0 in the other.

Separation of the marginals and the L\'evy copula is based on Sklar's
theorem for L\'evy measures. Due to the fact that all \lp
es with the exception of a CPP have a
singularity in 0, the \lm\ is considered on quadrants in $\bbr^d$
avoiding the origin. The simplest object to consider is, hence, a
$d$-dimensional subordinator, which allows for only positive jumps
in all components. We restrict ourselves in this paper to such
processes, extensions to general L\'evy processes are not
difficult, but notationally involved; see Kallsen and
Tankov~\cite{KalTan06} or Eder and Kl\"uppelberg~\cite{EdeKlu08}.

Our paper is organised as follows. In Section~\ref{s2} we
introduce the notion of a \lc\ needed later to model the
dependence structure between the components of a multivariate \lp.
Here we also explain the truncation scheme of the observed jumps
and present our prominent example, the bivariate $\al$-stable
Clayton subordinator. Section~\ref{s4} is dedicated to the
two-step estimation procedure. We prove consistency and asymptotic normality of
the IFM estimates in Section~\ref{s5} including the calculation of the
covariance matrix as the inverse of the Godambe information
matrix. For a comparison with the MLE based
on the full model we calculate its log-likelihood function in
Section~\ref{s7}. Finally, in Section~\ref{s8}, we perform a small
simulation study, where we compare the quality of all three
estimation methods: the full MLE, the full MLE based on joint
jumps only, and the estimates from the two-step procedure.

\section{Preliminaries}\label{s2}

\subsubsection*{The \lc}

Throughout this paper we denote by
$\bS=\left(\bS(t)\right)_{t\ge0}$ an increasing \lp\ with values
in $\bbr_+^d$ defined on a filtered probability space
$(\Omega,\calf,(\calf_t)_{t\ge0},\bbp)$.
This means that $\bS$ is a subordinator without Gaussian component, drift $\ga$
and a \lm\ $\Pi$ on $\bbr_+^d$ satisfying $\Pi(\{\bzero\})=0$ and
$\int_{\bbr_+^d}\min\{x,1\} \Pi(dx)<\infty$; cf. Sato~\cite{Sato:1999}, Th. 21.5, or Cont and Tankov~\cite{ContTankov}, Prop.~3.10.

The tail integral of the \lm\ $\Pi$ is the function
$\ov\Pi:[0,\infty]^d\rightarrow[0,\infty]$ defined by
 \beam\label{ti}
\ov\Pi(x_1,\ldots,x_d)= \left\{\begin{array} {ll}
\Pi([x_1,\infty)\times\cdots\times [x_d,\infty))\,,&
(x_1,\ldots,x_d)\in [0,\infty)^d\setminus\{\bzero\}\,\\
0\,, &  x_i=\infty \mbox{ for at least one $i$},\\
\infty\,,& (x_1,\ldots,x_d)={\bzero}.
\end{array}\right.
\eeam
The marginal tail integrals are defined analogously for
$i=1,\ldots,d$ as $\ov\Pi_i(x)=\Pi_i([x,\infty))$ for $x\ge 0$;
cf. Cont and Tankov~\cite{ContTankov}, Def.~5.7, and Kallsen and
Tankov~\cite{KalTan06}, Def.~3.3 and~3.4.

The jump dependence of the process $\bS$ is part of the multivariate tail integral and can be described by a so-called \lc.
We recall the notion of a \lc\ from \cite{ContTankov,KalTan06} to be a measure defining function
$\fC:[0, \infty]^d \to [0,\infty]$ with Lebesgue margins $\fC_k(u)=u$ for all $u\in[0,\infty]$ and $k=1,\ldots,d$.

The following result, called Sklar's Theorem for \lc s, is central
for our set-up; it has been proved in Cont and Tankov~\cite{ContTankov}, Th.~5.4, for a bivariate \lp\ and in Kallsen and
Tankov~\cite{KalTan06}, Th.~3.6, for a $d$-dimensional \lp.

 \bthe Let
$\ov\Pi$ denote the tail integral of a spectrally positive
$d$-dimensional L\'evy process, whose components have L\'evy
measures $\Pi_1,\ldots,\Pi_d$. Then there exists a L\'evy copula
$\fC:[0, \infty]^d \to [0,\infty]$ such that for all $x_1,
x_2,\ldots,x_d \in [0, \infty]$ \beam\label{cop} \ov\Pi(x_1,
\ldots, x_d)=
\fC\left({\ov\Pi_1(x_1)},\ldots,{\ov\Pi_d(x_d)}\right). \eeam If
the marginal tail integrals are continuous, then this L\'evy
copula is unique.
Otherwise, it is unique on $Ran (\ov\Pi_1)\times\cdots\times Ran (\ov\Pi_d)$.\\
Conversely, if $\fC$ is a L\'evy copula and $
\ov\Pi_1,\ldots,\ov\Pi_d$ are one-dimensional tail integrals
of spectrally positive L\'evy processes, then the relation
\eqref{cop} defines the tail integral of a $d$-dimensional
spectrally positive L\'evy process and $\ov\Pi_1,\ldots,\ov\Pi_d$
are tail integrals of its components. \ethe

\subsubsection*{Truncation of the small jumps}\label{s3}

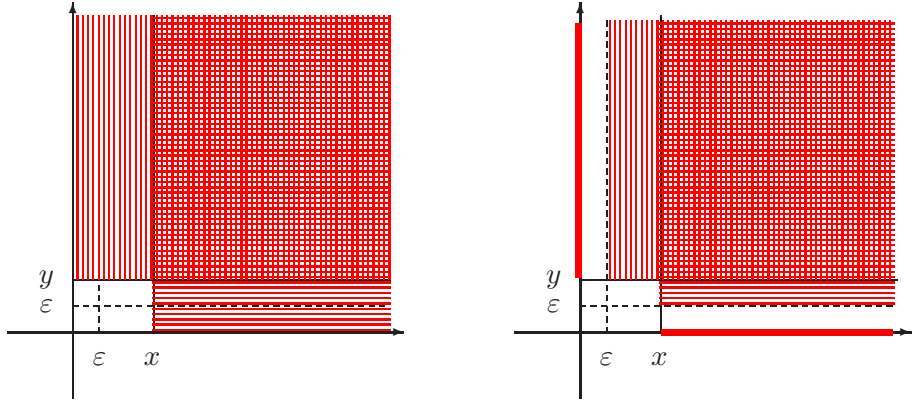
\begin{figure}[t]
\begin{center}
$\barr{ccc}
\begin{picture}(150,150)
 \put(0,25){\vector(1,0){150}} 
 \put(25,0){\vector(0,1){150}} 
 \multiput(35,25)(0,4){30}{\line(0,1){2}}
 \multiput(25,35)(4,0){30}{\line(1,0){2}}
 \put(15,35){\makebox(0,0){$\eps$}}  
 \put(35,15){\makebox(0,0){$\eps$}}  
 \put(25,45){\line(1,0){120}}        
 \put(55.5,25){\line(0,1){120}}      
 \put(55,15){\makebox(0,0){$x$}}     
 \put(15,45){\makebox(0,0){$y$}}     
 \h{\multiput(55,26)(0,2){60}{\line(1,0){90}}}     
 \h{\multiput(23,45)(2,0){60}{\line(0,1){100}}}  
 \end{picture} &$\qquad$&
 \begin{picture}(150,150)
 \put(0,25){\vector(1,0){150}} 
 \put(25,0){\vector(0,1){150}} 
 \multiput(35,25)(0,4){30}{\line(0,1){2}}
 \multiput(25,35)(4,0){30}{\line(1,0){2}}
 \put(15,35){\makebox(0,0){$\eps$}}  
 \put(35,15){\makebox(0,0){$\eps$}}  
 \put(25,45){\line(1,0){120}}        
 \put(55.5,25){\line(0,1){120}}      
 \put(55,15){\makebox(0,0){$x$}}     
 \put(15,45){\makebox(0,0){$y$}}     
 \h{\multiput(55,36)(0,2){54}{\line(1,0){89}}}     
 \h{\multiput(32.5,45)(2,0){54}{\line(0,1){98}}}  
 \h{\multiput(16,46)(0,0.7){138}{\line(1,0){2}}}     
 \h{\multiput(45,24)(0.7,0){125}{\line(0,1){2}}}  
 \end{picture}
       \earr$
\end{center}
 {\caption{{\label{pro-mass}}
Illustration of the tail integral $\ov\Pi$ of a truncated
bivariate \lp\ at $(x,y)$ for a process with jump sizes of
$\max\{x,y\}\ge\eps$ (left) and a process with jump sizes of
$x\ge\eps$ and $y\ge\eps$ (right). Note that in the right plot the
mass, where only one component is larger than $\eps>0$ and the other
smaller, is projected to the axes.}}
\end{figure}

For notational convenience we proceed with a bivariate subordinator.
As truncation point we choose $\eps>0$.
Figure~\ref{pro-mass} shows how the \lm\ $\Pi$ on $\bbr_+^2\setminus(0,\eps)^2$ is divided into two parts, the part concentrated on
$[\eps,\infty)^2$, and the part concentrated on the axes,
 which is in fact the projected measure of $\Pi$ on $[\eps,\infty)\times(0,\eps)$ and $(0,\eps)\times [\eps,\infty)$ to the
 horizontal and vertical axes, respectively.

\subsubsection*{The observation scheme}

It is based on all jumps of the process larger
than some $\eps$ componentwise  within the observation interval
$[0,t]$. That is, we may observe a single jump $x$ or $y$ either
in the first or in the second component. The other observed jumps
are $(x,y)$, where $x\ge\eps$ and $y\ge\eps$ at the same time. Let
$n=n_1+n_2$ denote the total number of jumps occurring in $[0,t]$
in either component, where we denote by $n_1$ and $n_2$ the number
of jumps in each marginal process, respectively. This means that
we count every joint jump in both components as two jumps. Then
$n$ decomposes in the number $n_1^\bot$ of jumps occurring only in
the first component, the number $n_2^\bot$ of jumps occurring only
in the second component, and the number $2n^{\|}$ of jumps
occurring in both components.

We denote by $(x_1,\ldots,x_{n_1},y_1,\ldots,y_{n_2})$ the
observed jumps. By the independence property of the jumps of a
L\'evy process the order does not matter as long as concurrent jumps remain in the same coordinate. Consequently, throughout
the paper we place  w.l.o.g. all joint jumps at the beginning of
the $x$- and $y$-observations so that
$(\bx^\|,\by^\|)=\left((x_1,y_1),\ldots,(x_{n^\|},y_{n^\|})\right)$.

The resulting $n^{\|}+n_1^\bot+n_2^\bot$ observations can be attributed to a
bivariate CPP similar to the model considered in Esmaeili and
Kl\"uppelberg~\cite{EsK1}. We shall need the marginal truncated
L\'evy measures $\Pi_k^{(\eps)}$ for $k=1,2$. They will be
calculated by first determining the L\'evy measures of those
processes representing joint jumps larger than $\eps$, denoted by
{$\Pi^{(\eps)\|}$}, single jumps larger than $\eps$ in the first
or second component, denoted by {$\Pi_1^{(\eps)\bot}$} and
{$\Pi_2^{(\eps)\bot}$}, respectively.

The tail integrals of the observed CPP are given for $x,y>\eps$ by
\beam\label{tails}
 {\ov\Pi}^{(\eps) \|}(x,y)\,&=& \,\ov\Pi(x,y)\,,\nonumber\\
  {\ov\Pi_1}^{(\eps)\bot}( x) \,&=& \, \ov\Pi( x,0)-\ov\Pi(
  x,\eps)\,,\\
 {\ov\Pi_2}^{(\eps)\bot}(y) \, &=& \,
 \ov\Pi(0,y)-\ov\Pi(\eps,y)\,.\nonumber
\eeam
The jump intensities of these CPPs are
  \beam\label{intensities}
 {\la}^{(\eps)\|} &=&
 \ov\Pi(\eps,\eps)\,, \nonumber\\
 {\la_1}^{(\eps)\bot} &=& \ov\Pi(\eps,0)-\ov\Pi(\eps,\eps)\,,\\
 {\la_2}^{(\eps)\bot} &=&
 \ov\Pi(0,\eps)-\ov\Pi(\eps,\eps).\nonumber
 \eeam
The corresponding jump size distributions are given by the \lm s
divided by the intensities, respectively.
The marginal tail integrals of the truncated processes are now calculated as
 \beao
 \ov\Pi_1^{(\eps)}(x) &=& \ov\Pi^{(\eps)\|}(x,\eps)
+\ov\Pi_1^{(\eps)\bot}(x) =\ov\Pi(x,0)\,,\quad x\ge\eps\,\\
\ov\Pi_2^{(\eps)}(y) &=& \ov\Pi^{(\eps)\|}(\eps,y)
+\ov\Pi_2^{(\eps)\bot}(y) =\ov\Pi(0,y)\,,\quad y\ge\eps\,,
 \eeao
which implies intensities $\la_k^{(\eps)}=
\la^{(\eps)\|}+{\la_k^{(\eps)\bot}}=\ov\Pi_k(\eps)$.

Lemma 4.1 in Esmaeili and Kl\"uppelberg~\cite{EsK2} explains the
consequence of the small jumps truncation to the \lc. We shall
need the notion of a generalized inverse function: for
$g:\bbr\to\bbr$ increasing define the {\em generalized inverse} of
$g$ as $g^{\leftarrow}(x)=\inf\{u\in\bbr : g(u)\ge x\}$. The
definition extends naturally to other supports. For more details
and properties of the generalized inverse we refer to
Resnick~\cite{resnick:1987}, Section~0.2.

 From Lemma 4.1 in Esmaeili and Kl\"uppelberg~\cite{EsK2} the \lc\
 of the CPP  is given by
 \beam\label{C-tilde}
 \fC^{(\eps)}(u,v) =\fC\left(\fC^\leftarrow_1({u},{\leps_{2}}),
\fC^\leftarrow_2({\leps_{1}},{v})\right),\quad 0<u,v< {\lepsd},
\eeam where for $k=1,2$ the symbol $\fC^\leftarrow_k$ denotes  the
generalized inverse of $\fC$ with respect to
 the $k$-th argument.

The following will be our prominent example.

\bexam[Bivariate $\al$-stable Clayton subordinator]\label{bss}\\
Let $c_1,c_2>0$ and $0<\al_1,\al_2<1.$ Assume that
$\ov\Pi_1(x)=c_1x^{-\al_1}$ for $x>0$ and
$\ov\Pi_2(y)=c_2y^{-\al_2}$ for $y>0$ and that dependence is
modeled by a Clayton  \lc, which is given by
$$\fC(u,v)=\left(u^{-\delta}+v^{-\delta}\right)^{-1/\delta}\,,\quad u,v>0\,,$$
with dependence parameter $\delta>0$.

By \eqref{tails} the tail integrals of the observed CPP are given by
\beam\label{jointLevy}
{\ovPieps}^\|(x,y)&=& \left((c_1x^{-\al_1})^{-\de} +(c_2y^{-\al_2})^{-\de}\right)^{-\frac{1}{\de}}\,,\quad x,y\ge\eps\,, \\
 {\ovPieps_1}^\bot( x)&=&
 c_1 x^{-\al_1}\left[1-\left(1+\Big(\frac{c_2\eps^{-\al_2}}{c_1 x^{-\al_1}}\Big)^{-\de}\right)^{-1/\de}\right],\quad x\ge\eps\,,\nonumber\\
 {\ovPieps_2}^\bot(y)&=&
 c_2y^{-\al_2}\left[1-\left(1+\Big(\frac{c_1\eps^{-\al_1}}{c_2 y^{-\al_2}} \Big)^{-\de}\right)^{-1/\de}\right],\quad y\ge\eps\,.\nonumber
\eeam
From \eqref{intensities} we calculate the jump intensities
\beam\label{jointint}
 {\lepsd} &=& \left((c_1\eps^{-\al_1})^{-\de} +{(c_2\eps^{-\al_2})^{-\de}}\right)^{-\frac{1}{\de}},\\
 {\leps_1}^\bot &=&
 c_1 \eps^{-\al_1}\left[1-\left(1+\Big(\frac{c_2\eps^{-\al_2}}{c_1 \eps^{-\al_1}}\Big)^{-\de}\right)^{-1/\de}\right],\nonumber\\
 {\leps_2}^\bot&=&
 c_2\eps^{-\al_2}\left[1-\left(1+\Big(\frac{c_1\eps^{-\al_1}}{c_2 \eps^{-\al_2}} \Big)^{-\de}\right)^{-1/\de}\right].\nonumber
 \eeam
 The marginal tail integrals and intensities of the truncated process are now calculated
for $k=1,2$ as
$$\ov\Pi_k^{(\eps)}(x)=c_k x^{-\al_k}\,,\quad x\ge\eps\,,\quad\mbox{and}\quad \la_k^{(\eps)}=c_k\eps^{-\al_k}\,.$$
This implies for the marginal jump size distributions
 \beao
 P(X>x)&=&  \ov\Pi_1^{(\eps)}(x)/\la_1^{(\eps)} = \eps^{\al_1} x^{-\al_1}\,,\quad x\ge \eps\,,\\
 P(Y>y)&=&  \ov\Pi_2^{(\eps)}(y)/\la_2^{(\eps)} = \eps^{\al_2} y^{-\al_2}\,,\quad y\ge \eps\,.
 \eeao
By \eqref{C-tilde} the \lc\
of the observed CPP is for $0<u,v<\la^{(\eps)\|}$ given by
\beao
\fC^{(\eps)}(u,v)
&=&\fC\left(\left(u^{-\de}-({\leps_{2}})^{-\de}\right)^{-1/\de}, \left(v^{-\de}-({\leps_{1}})^{-\de}\right)^{-1/\de}\right)\\
&=&\left(u^{-\de}+v^{-\de}-(c_1^{-\de}\eps^{\al_1\de}+
c_2^{-\de}\eps^{\al_2\de})\right)^{-1/\de}\,.
\eeao
\eexam

\section{Two-step parameter estimation of a \lp}\label{s4}

The idea of a two-step procedure for subordinators is similar
to the IFM method for multivariate distributions. The term IFM is
the acronym for ``inference functions for margins" and has been
applied in various areas of multivariate statistics; cf. Godambe~\cite{God91} and Joe~\cite{Joe:MultiModelsDepConcepts}, Ch.~10.
Obviously, the maximization of a likelihood with many parameters can
be numerically sophisticated and computationally time-consuming;
in a two-step method the parameters of the marginal
components are estimated first and the L\'evy copula parameters in a second
step, thus reducing the dimensionality of the problem.
For multivariate distribution functions, the algorithm is
explained, for instance, in Joe~\cite{Joe:MultiModelsDepConcepts},
Ch.~10.

For a multivariate \lp\ in $\bbr^d$ for arbitrary dimension $d\in\bbn$, the {two-step} algorithm can be
formalized as follows.

{\bf Step 1 : } We do not distinguish between single and common
jumps, but make use of all data available; i.e., we take all
observations $x_{ik}>\eps$ for $i=1,\ldots,n_k$ and all
$k=1,\ldots,d$. We denote by $\ga=(\thet_1,\ldots,\thet_d)$ the
vector of all marginal parameters (the $\theta_i$ are usually
vectors) and let $l_1^{(\eps)},\ldots,l_d^{(\eps)}$ be the
marginal log-likelihood functions with respect to the parameters.
Determine \beam\label{step1} \wt\ga := \argmax_{\ga} \sum_{k=1}^d\
l_k^{(\eps)}(\thet_k\mid \bx_{k})\,, \eeam where
$\bx_k=(x_{1k},x_{2k},\ldots,x_{n_k k})$ are all observations in
component $k$ larger than $\eps$.

{\bf Step 2 : } Write the log-likelihood $l^{(\eps)}$ of a CPP, whose jumps are only the common jumps
of $x_{ik}>\eps$ for $i=1,\ldots,n^\|$ and $k=1,\ldots,d$, plug in
the marginal parameter estimates from Step 1, resulting in the
log-likelihood of a CPP with only dependence
structure parameter $\de$. Maximize the log-likelihood function
over $\de$; i.e., estimate the \lc\ parameter vector
$\delta\in\bbr^m$ for some $m\in\bbn$, based on the common jumps:
\beam\label{step2} \wt\delta := \argmax_{\delta}\,
l^{(\eps)}(\delta\mid\wt\ga,\bx_1^\|,\ldots,\bx_d^\|)\,, \eeam
 where $\wt\ga=(\wt\thet_1,\ldots,\wt\thet_d)$ and $\bx_k^\|=(x_{1k},\ldots,x_{n^{\|}k})$ for $k=1,\ldots,d$.

\brem\label{sc}
The MLE $\hat\eta$
of the parameter vector $\eta=(\theta_1,\ldots,\theta_d,\de)$ is
derived by maximization of the log-likelihood of the multivariate
CPP $l^{(\eps)}$ over the parameter vector $\eta$ (as done in \cite{EsK2}).
The estimate $\hat\eta$ is the solution of
$$\left(\frac{\p l^{(\eps)}}{\p\thet_1},\ldots,\frac{\p l^{(\eps)}}{\p \thet_d},\frac{\p l^{(\eps)}}{\p\de} \right)=0.$$
 This is in contrast with the two-step method, where
the estimate $\wt\eta$ is the solution of
$$\left(\frac{\p l_1^{(\eps)}}{\p\thet_1},\ldots,\frac{\p l_k^{(\eps)}}{\p \thet_k},\frac{\p l^{(\eps)}}{\p\de} \right)=0.$$
\erem

\brem
The aim of the two-step method is in fact to
reduce the dimension of the parameter vector to have a
simpler structure for the optimization of the likelihood function.
Note that the observation scheme in \cite{EsK2}, which takes only
the $n^\|$ observations of the joint jumps in both steps into account,
fails this goal, since the observation scheme used there introduces the dependence parameter into the marginal likelihoods.
\erem

\subsection{Two-step estimation method of an $\al$-stable Clayton subordinator with different marginal parameters}\label{s41}

The following algorithm works in principle in every dimension. For notational simplicity we formulate it only for dimension $d=2$.
Let $\bS=(S_1,S_2)$ be a bivariate $\al$-stable Clayton
subordinator as introduced in Example~\ref{bss} with different
marginal parameters $\theta_1=(\al_1,c_1)$ and
$\theta_2=(\al_2,c_2)$ with $\al_k\in(0,1)$ and $c_k\in(0,\infty)$
for $k=1,2$ and a \lc\ parameter $\de\in(0,\infty)$. We assume the
observation scheme as described in Section~\ref{s3}. We denote by
$(X_1,\ldots,X_{n^\|},\ldots,X_{n_1},Y_1,\ldots,Y_{n^\|},\ldots,Y_{n_2})$
the vector of jumps
larger than $\eps$ for the component processes $S_1^{(\eps)}$ and
$S_2^{(\eps)}$, respectively. As before, all double jumps are numbered as
$(X_i,Y_i)$ for $i=1,\ldots,n^\|$.

{\bf Step 1 : }
Since the marginal log-likelihoods have the same
structure with no common parameters, \eqref{step1} decomposes in its components for $S_1$ and $S_2$,
and maximization is done separately.
We proceed as in Basawa and Brockwell~\cite{BaB1,BaB2}; cf. Esmaeili and Kl\"uppelberg~\cite{EsK2}, Example~3.1,
and exemplify it for the first component:
\beao
 l^{(\eps)}_1(\log c_1,\al_1; \bx)=-c_1 t\eps^{-\al_1}+n_1({\log c_1} +\log\al_1)-(\al_1+1)\sum_{i=1}^{n_1}\log x_{i}\,.
 \eeao
From Basawa and Brockwell~\cite{BaB1,BaB2} we know that the
marginal MLEs of $c_1$ and $\al_1$ and the intensity parameter
$\leps_1$ are given by
\beam\label{margins}
\wt\la^{(\eps)}_1 &=& \frac{n_1}{t}\,,\nonumber\\
\wt\al_1 &=& \left(\frac{1}{n_1}\sum_{i=1}^{n_1}\Big(\log
X_{i}-\log\eps\Big)
+\log\eps\Big(1-\frac{\la^{(\eps)}_1}{\wt\la^{(\eps)}_1}\Big)
\right)^{-1}\,,\\
\log \wt c_1 &=& \log \wt\la^{(\eps)}_1 + \wt\al_1\log\eps\,.\nonumber
 \eeam
Furthermore, asymptotic normality holds with degenerate limit for
$(\wt c_1, \wt \al_1)$ and with asymptotic independence for
$(\wt\la_1, \wt \al_1)$  as $n_1\to\infty$.
Limit laws hold for both situations, $t\to\infty$ or  $\eps\to0$.
The first limit was
derived in Basawa and Brockwell~\cite{BaB2}. Asymptotic
independence for the second vector was shown in H\"opfner and
Jacod~\cite{HJ}. Both results are reported with precise  rates and
the asymptotic covariance matrix in Esmaeili and
Kl\"uppelberg~\cite{EsK2}, Example~3.1.

{\bf Step 2 : } We first determine the log-likelihood function in
\eqref{step2} for the bivariate CPP of common jumps larger than
$\eps$. By
\eqref{jointint} the intensity is
$\la^{(\eps)\|}=(c_1^{-\de}\eps^{\al_1\de}+c_2^{-\de}\eps^{\al_2\de})^{-\frac1{\de}}$.
Together with \eqref{jointLevy} this yields the survival function
of bivariate joint jumps
\beao
 \ov F^{(\eps)}(x,y)=\left(\frac{c_1 ^{-\de}x^{\al_1\de}+
 c_2 ^{-\de}y^{\al_2\de}}{c_1 ^{-\de}\eps^{\al_1\de}+c_2^{-\de}\eps^{\al_2\de}}\right)^{-\frac1{\de}},\quad
 x,y\ge\eps,
 \eeao
with density given by
\beam \label{jointden1}
f^{(\eps)}(x,y) =\frac{\al_1\al_2(1+\delta)
(c_1^{-\de}\eps^{\al_1\de}+c_2^{-\de}\eps^{\al_2\de})^{\frac{1}{\delta}}} {(c_1c_2)^{\de}}
\frac{x^{\al_1\delta-1}y^{\al_2\delta-1}}
{(c_1^{-\de}x^{\al_1\delta}+c_2^{-\de}y^{\al_2\delta})^{\frac{1}{\delta}+2}}\,.
\eeam
This results in the log-likelihood function
 \beao
 \lefteqn{l^{(\eps)}(c_1,c_2,\al_1,\al_2,\de;{\bx}^\|,{\by}^\|) \, = \, {-\la^{(\eps)\|} t}+
n^{\|} \log(1+\de)-n^{\|}\de(\log c_1+\log c_2)}\\
&& + n^{\|}(\log\al_1 +\log\al_2)
 +(\al_1\de-1)\sum_{i=1}^{n^{\|}}\log x_i+
(\al_2\de-1)\sum_{i=1}^{n^{\|}}\log
y_i\\
&&-(\frac1{\de}+2)\sum_{i=1}^{n^{\|}}\log
\left(c_1^{-\de}x_i^{\al_1\de}+c_2^{-\de}y_i^{\al_2\de}\right),\nonumber
\eeao
where $(\bx^\|,\by^\|)=((x_1,y_1),\ldots,(x_{n^\|},y_{n^\|}))$.

Given the marginal parameter estimates from the first step, the score function with respect to the dependence parameter
 $\de$ is given by
\beao
\lefteqn{\frac{\partial
l^{(\eps)}(\delta\mid\wt\ga,\bx^\|,\by^\|)}{\partial\delta} \, = \,
-\frac{\p\lepsd}{\p\de}
t+\frac{n^{\|}}{1+\delta}-n^\|(\log\wt c_1+\log \wt
c_2)}\\
&& +\wt\al_1\sum_{i=1}^{n^{\|}} \log x_i +\wt\al_2\sum_{i=1}^{n^\|}\log y_i +\frac1{\delta^2}
\sum_{i=1}^{n^{\|}}\log\Big(\wt c_1^{\
-\de}x_i^{\wt\al_1\delta}+\wt c_2^{\
-\de}y_i^{\wt\al_2\delta}\Big)\\
&&-\Big(\frac1{\delta}+2\Big)\sum_{i=1}^{n^{\|}}
 \frac{\partial}{\partial\delta}
  \log\Big(\wt c_1^{\ -\de}x_i^{\wt\al_1\delta}+
{\wt c_2}^{\ -\de}y_i^{\wt \al_2\delta}\Big). \eeao
 The parameter estimate $\wt\de$ can be found numerically by solving the following equation for $\de$:
  $$ \frac{\partial
l^{(\eps)}(\delta\mid \wt\ga,\bx^\|,\by^\|)}{\partial\delta}=0.$$

\brem\label{sc2}
The vector of score functions in the two-step method is given by
\beao
\bJ^{(\eps)}(\mathbf{X,Y};\eta) \, = \, \hspace*{12cm}\\
\Big(\frac{\partial l_1^{(\eps)}(\log
c_1,\al_1;\mathbf{X})}{\p\log c_1}, \frac{\partial
l_1^{(\eps)}(\log c_1,\al_1;\mathbf{X})}{\p \al_1}, \frac{\p
l_2^{(\eps)}(\log c_2,\al_2;\mathbf{Y})}{\partial \log c_2},
\frac{\p l_2^{(\eps)}(\log c_2,\al_2;\mathbf{Y})}{\partial \al_2},
\frac{\partial
l^{(\eps)}(\delta;\mathbf{X}^\|,\mathbf{Y}^\|)}{\partial
\delta}\Big)^T, \eeao where $\eta=(\log c_1,\log
c_2,\al_1,\al_2,\de)^T$ is the parameter vector,
$\mathbf{X}=(X_1,\ldots,X_{n_1})$,
$\mathbf{Y}=(Y_1,\ldots,Y_{n_2})$ and
$(\mathbf{X^\|,Y^\|})=((X_1,Y_1),\ldots,(X_{n^\|},Y_{n^\|}))$.
\erem

\subsection{Two-step method for a bivariate $\al$-stable Clayton subordinator with common marginal
parameters}\label{s42}

For an analysis of the two-step estimation procedure we simplify
the model as follows.
 Let $\bS=(S_1,S_2)$ be  a bivariate $\al$-stable subordinator as
 in Example~\ref{bss} with common marginal parameters
 $\thet_1=\thet_2=(\al,c)$ and a Clayton \lc\ parameter $\de$.
 Assume  an observation scheme as explained in Section~\ref{s3}.
 Maximum likelihood estimation for the parameters of this model
 was discussed in Esmaeili and Kl\"uppelberg~\cite{EsK2} in detail.
 In this section we estimate the parameters with the two-step
 method.

{\bf Step 1 : }
The  log-likelihood function \eqref{step1}, which ignores the dependence structure, is given by
 \beam\label{like1}
 l_{12}^{(\eps)}(\log c,\al)&=&l_{1}^{(\eps)}(\log c,\al)+l_{2}^{(\eps)}(\log c,\al)\nonumber\\
 &=& -2ct\eps^{-\al}+ n (\log
 c+\log\al)-(\al+1)\sum_{i=1}^{n}\log z_{i},
 \eeam
 where $n:=n_1+n_2$ is Poisson distributed.
 Since $n_1$ and $n_2$ have both intensity $\leps:=\leps_1=\leps_2$, $n$ has intensity $2\leps=2c\eps^{-\al}$
 and $(z_1,\ldots,z_n)=(x_1,\ldots,x_{n_1},y_1,\ldots,y_{n_2})$.
  Note that the corresponding random variables
 $\log(\frac{Z_i}{\eps})$, for $i=1,\ldots,n$ are exponentially
 distributed  with density $f(u)=\al e^{-\al u}$ for  $u>0$.
The log-likelihood has score functions with respect to the marginal parameters  $\log c$  and $\al$
as follows:
 \beam\label{score1}
 \frac{\p l_{12}^{(\eps)}(\log c,\al)}{\p\log c}
 &=& n-2 c t\eps^{-\al} = n-2\leps t\\
 \frac{\p
 l_{12}^{(\eps)}(\log c,\al)}{\p\al}
 &=&\frac{n}{\al}+2ct\eps^{-\al}\log\eps-\sum_{i=1}^{n}\log  z_i
 = -\sum_{i=1}^{n} \Big(\log  \frac{z_i}{\eps}-\frac1{\al}\Big) - (n-2\leps t) \log\eps \nonumber
\,.
 \eeam
The common intensity parameter $\la^{(\eps)}=c\eps^{-\al}$ and the
marginal parameters $\log c$ and $\al$ can be estimated by
\eqref{margins} as \beao
 \wt\la^{(\eps)}&=&\frac{n}{2t}\\
 \wt\al&=&\bigg(\frac1{n}\sum_{i=1}^{n}(\log Z_i-\log\eps) +\big(1-\frac{\leps}{\wt\la^{(\eps)}}
 \big) \log\eps \bigg)^{-1}\\
 \log \wt c&=& \log\wt\la^{(\eps)}+ {\wt \al}\log\eps\,.
 \eeao

 {\bf Step 2 : }
 As explained in Esmaeili and Kl\"uppelberg~\cite{EsK2}, for simplifying the calculations of the second derivatives later we reparameterize the dependence to  $\thet=\al \de$.
 The joint density of bivariate jumps is a special case of
 \eqref{jointden1} and has been calculated in (4.10) in Esmaeili
 and Kl\"uppelberg~\cite{EsK2}.
 From \eqref{jointint} we know that ${\lepsd}=c\eps^{-\al}2^{-\frac{\al}{\theta}}$,
 which we use for abbreviation.
Then the log-likelihood in \eqref{step2} is
 \beam\label{like2}
 l^{(\eps)}(\log c, \al,\thet) &=&  - {\lepsd}  t +  n^\|\log\al + n^\|\log(\al+\theta)+ n^\|\log c
 \\
 &&
 +(\theta -1)\sum_{i=1}^{n^\|} (\log x_i+\log y_i)
 -(2+\frac{\al}{\theta})\sum_{i=1}^{n^\|}
 \log(x_i^{\theta}+y_i^{\theta})\nonumber \,.
 \eeam
 The score function with respect to the parameter $\theta$ is then
 given by (the derivatives of $\lepsd$ are calculated in Lemma~\ref{lepsdder} below)
 \beam\label{score2}
 \frac{\p l^{(\eps)}}{\p \thet}&=& - \frac{\p{\lepsd}}{\p\theta} t
 +\frac{n^\|}{\al+\thet}
 + \sum_{i=1}^{n^\|}(\log x_i+\log
 y_i)\\
 &&+\frac{\al}{\thet^2}\sum_{i=1}^{n^\|}\log(x_i^\thet+y_i^\thet)
 -(2+\frac{\al}{\thet})
 \sum_{i=1}^{n^\|}\frac{\p}{\p\thet}\log(x_i^\thet+y_i^\thet).\nonumber
 \eeam
 Given the estimates of the marginal parameters $\wt c$ and $\wt \al$ from the
 first step, the estimate of $\theta$ can be computed numerically as the argmax of the right hand side of
 \eqref{score2}.

 \brem\label{rem2}
 The vector of score functions from Remark~\ref{sc2} reduces to
 \beam\label{score-func2}
 \bJ^{(\eps)}(\mathbf{X,Y};\eta)= \Big(\frac{\partial l_{12}^{(\eps)}(\log c,\al;\mathbf{Z})}{\p \log c},
 \frac{\p l_{12}^{(\eps)}(\log c,\al;\mathbf{Z})}{\partial \al},
 \frac{\partial l^{(\eps)}(\log c,\al,\theta;\mathbf{X^\|,Y^\|})}{\partial
 \thet}\Big)^T,
 \eeam
 where $\eta=(\log c,\al,\thet)^T$ is the
 parameter vector, $\mathbf{Z}=(X_1,\ldots,X_{n_1},Y_1,\ldots,Y_{n_2})$ and
 $(\mathbf{X^\|},\mathbf{Y^\|})=(X_1,Y_1),\ldots,(X_{n^\|},Y_{n^\|})$.
\erem

\section{Asymptotic properties of the two-step estimates}\label{s5}

The two-step estimation procedure is a special case of the estimating functions approach, which goes back to Godambe~\cite{God91} (see also the Z estimates in van der Vaart~\cite{Vaart}).
In this framework, the Godambe information matrix plays the role of the Fisher information matrix in classical MLE.

We explain this for L\'evy copulas.
Let $\bS=(S_1,S_2)$ be a bivariate $\al$-stable Clayton
subordinator with parameter vector $\eta\in\bbr^k$ including
marginal and dependence parameters.
Assume further an observation
scheme as explained in Section~\ref{s3}.
In principle the two-step estimation procedure can be applied to both situations of
Section~\ref{s41} with $\eta\in\bbr^5$ or of Section \ref{s42} with $\eta\in\bbr^3$.

For the vector of score functions, denoted by
$\bJ^{(\eps)}(\mathbf{X,Y};\eta)$ as in Remarks~\ref{sc2}
and~\ref{rem2}, the so-called {\em Godambe information matrix} is
calculated for fixed $\eps>0$ as \beam\label{godambe} G &:=&
D^{\top}M^{-1}D, \eeam
 where
\beam
D &:=&\frac1{2\leps
t}\Bbb{E}\Big[-\frac{\p\bJ^{(\eps)}(\mathbf{X,Y};\eta)}{\p
\eta}\Big],\label{dgoda}\\
M&:=&\frac1{2\leps t}\Bbb{E}\Big[\bJ^{(\eps)}(\mathbf{X,Y};\eta)
{\bJ^{(\eps)}(\mathbf{X,Y};\eta)}^T\Big] \label{mgoda}
\eeam
are $k\times k$-matrices.
Under appropriate conditions, which will be shown below, the
asymptotic covariance matrix of $n^{-\frac1{2}}(\wt\eta-\eta)$ is
equal to the inverse of $G$.

For the remainder of this section we restrict the process $\bS$ again
to the model in Section~\ref{s42},  a bivariate $\al$-stable
subordinator with common marginal parameters $\log c$ and
$\al$, and dependence parameter $\thet$.
We denote by $l^{(\eps)}_{12}$ the log-likelihood
of the common marginal parameters $\ga:=(\log c, \al)$ as in
\eqref{like1}, and by $l^{(\eps)}$ the log-likelihood of the
bivariate CPP in the second step as in
\eqref{like2}. Assume further that $\eta_0=(\log
c_0,\al_0,\thet_0)$ is the true parameter vector. We prove
consistency of the two-step estimators, and their joint asymptotic
normality. We calculate the Godambe information matrix $G$ as well
as the asymptotic covariance matrix of the estimators.

There is a fundamental difference between our approach and the classical used for distributional copulas in Joe~\cite{Joe:MultiModelsDepConcepts}, Section~10.1.1. Whereas he can work with
a fixed number of multivariate data, our process structure with observations on an interval $[0,t]$ implies a random number of data points. Moreover, we have to deal with the problem of single and common jumps.
Furthermore, \cite{Joe:MultiModelsDepConcepts}  assumes regularity conditions like interchangeability of derivatives and integrals, which are not necessarily guaranteed in our context (cf. Esmaeili and Kl\"uppelberg~\cite{EsK2}, Section~4.2). As a consequence, we will provide a full proof of the asymptotic normality of the IFM estimators in Theorem~\ref{asnor} below.

\subsection{Auxiliary results}\label{s51}

We shall need the following derivatives of $\lepsd$.

\ble\label{lepsdder}
For $\lepsd=c\eps^{-\al} 2^{-\frac{\al}{\theta}}$ the partial derivatives are given by
\beao
\frac{\p{\lepsd}}{\p\log c} &=& \lepsd\,,\quad
\frac{\p{\lepsd}}{\p\al} \, = \,  -\lepsd \big(\log\eps +\frac1{\theta}\log 2\big)\,,\quad
\frac{\p{\lepsd}}{\p\theta} \, = \,  \lepsd\frac{\al\log2}{\thet^2}.
\eeao
The second derivatives can be calculated as
\beao
\frac{\p^2{\lepsd}}{\p\theta\p\log c} &=& \lepsd \frac{\al\log 2}{\theta^2}\,,\\
\frac{\p^2{\lepsd}}{\p\theta\p\al} &=& - \lepsd \frac{\log 2}{\theta^2}\Big(\al\log\eps + \frac{\al }{\theta}\log 2 -1\Big)\,,\\
\frac{\p^2{\lepsd}}{\p\theta^2} &=&  \lepsd\frac{\al\log
2}{\theta^2}\Big(\frac{\al\log 2}{\theta^2}-\frac{2}{\theta}\Big)\,.
\eeao
\halmos
\ele

We calculate several matrices for later use, where the details are given in the Appendix.
Throughout we abbreviate $d=\frac{\lepsd}{2\leps}=2^{-\frac{\al}{\theta}-1}$.

\ble\label{Heps}
We denote by $H^{(\eps)}=\frac{\p\bJ^{(\eps)}(\mathbf{X,Y};\eta)}{\p
\eta}$ the matrix of the second-order derivatives of
the two-step log-likelihood functions \eqref{like1} and \eqref{like2}, respectively.
Then
 \beam\label{H}
 H^{(\eps)}=  2\leps t \, \left(\barr{ccc}
 -1        &  \log\eps & 0\\
  \log\eps & -\frac{n }{\al^2 2\leps t}- (\log\eps)^2 & 0 \\
 - d  \,\frac{\al\log2}{\thet^2}&
  d  \big( \frac{\al\log2}{\thet^2}\log\eps-A \big) & - d  B
 \earr
 \right)
 \eeam
 where
\beao
A(\eta) &:=& {-}\frac{\al(\log2)^2}{\thet^3}+\frac{\log2
}{\thet^2}+\frac{n^\|}{\lepsd t(\al+\thet)^2}-\frac1{\lepsd t\thet^2}\sum_{i=1}^{n^\|}\log(X_i^\thet+Y_i^\thet)\\
&& + \frac1{\lepsd t\thet}\sum_{i=1}^{n^\|}\frac{\p}{\p\thet}\log(X_i^\thet+Y_i^\thet),\\
B(\eta) &:=&\Big(\frac{\al\log2}{\thet^2}\Big)^2-\frac{2\al\log2}{\thet^3}+\frac{n^\|}{\lepsd
t(\al+\thet)^2}
+\frac{2\al}{\lepsd t\thet^3}\sum_{i=1}^{n^\|}\log(X_i^\thet+Y_i^\thet)\\
&&-\frac{2\al}{\lepsd t\thet^2}
\sum_{i=1}^{n^\|}\frac{\p}{\p\thet}\log(X_i^\thet+Y_i^\thet)
+\frac{2\thet +\al}{\lepsd t\thet}
\sum_{i=1}^{n^\|}\frac{\p^2}{\p\thet^2}\log(X_i^\thet+Y_i^\thet).
\eeao
\halmos
\ele

We present the two matrices $D$ and $M$ from \eqref{dgoda} and \eqref{mgoda}, respectively.

 \ble\label{D}
 Assume a bivariate $\al$-stable Clayton subordinator with common marginal parameters $(\log c,\al)$ and dependence parameter $\theta=\al\delta$.
 Assume also the observation scheme given in Section~\ref{s2}.
  Recall that
  $\la^{(\eps)}=\leps_1=\leps_2=c\eps^{-\al}$ is the marginal intensity parameter,
  $\lepsd=c\eps^{-\al}2^{-\frac{\al}{\thet}}$ is the joint jumps intensity parameter.
Then the matrix $D=-\frac1{2\leps t} \bbe[ H^{(\eps)}]$ of \eqref{dgoda} is given by
 \beam\label{DstableC} D \, =
\left(\barr{ccc}
 1 &  -  \log\eps & 0\\
 - \log\eps &  \frac{1}{\al^2}+ (\log\eps)^2 & 0\\
d \frac{\al\log2}{\thet^2} & d
\big(- \frac{\al\log2}{\thet^2}\log\eps+ a \big) & d  \,b
 \earr
 \right),
\eeam
where
\beam
 a(\al,\theta) &=&
 -\frac{\al(\log2)^2}{\thet^3} + \frac{\log 2}{\theta^2} + \frac1{(\al+\theta)^2}
 - \frac1{\theta^2} \bbe\Big[\log(X_1^\thet+Y_1^\thet)\Big]
 +\frac1{\theta} \bbe\Big[ \frac{\p}{\p\theta}\log(X_1^\thet+Y_1^\thet)\Big]\nonumber\\
 \label{a}\\
  b(\al,\thet) & = & \Big(\frac{\al\log2}{\thet^2}\Big)^2-\frac{2\al\log2}{\thet^3}+\frac{1}{(\al+\thet)^2}
+\frac{2\al}{\thet^3}\bbe\Big[\log(X_1^\thet+Y_1^\thet)\Big]\nonumber\\
&&-\frac{2\al}{\thet^2}
\bbe\Big[\frac{\p}{\p\thet}\log(X_1^\thet+Y_1^\thet)\Big]
+\frac{2\thet +\al}{\thet}\bbe\Big[
\frac{\p^2}{\p\thet^2}\log(X_1^\thet+Y_1^\thet)\Big]. \label{b}
\eeam
\halmos
 \ele

 \brem\label{Dinv}
(i) \, The functions $a(\al,\theta)$ and $b(\al,\thet)$ are
deterministic functions of the parameters and do not depend on $t$
or $\eps$.
Moreover, all expectations in \eqref{a} and \eqref{b} are finite, since $\log X_i$ and $\log Y_i$ are exponentially distributed (cf. Lemma~4.4 of Esmaeili and Kl\"uppelberg~\cite{EsK2}.)\\
(ii) \,
 We shall need the following inverse, which exists for $b\neq 0$:
 {\beam\label{DstableCinv}
D^{-1} \, =
\left(\barr{ccc}
 1+\al^2(\log\eps)^2 &  \al^2\log\eps & 0\\
  \al^2 \log\eps &  \al^2 & 0\\
- \frac1{b} \big(a\al^2\log\eps +\frac{\al\log2}{\thet^2}\big) &
-\frac{a}{b}\al^2 &  \frac1{d\,b}
 \earr
 \right).
 \eeam}
\halmos
 \erem

In order to calculate the matrix $M$ from \eqref{mgoda} we shall
need the following result on the dependence of $n$ and $n^\|$.

 \ble\label{nnd}
Recall that $n=n_1+n_2=2n^\|+n^\bot_1+n^\bot_2$.
Then
\beao
 \Bbb E [n n^\|] &=& 2\lepsd t (1+\leps t)\quad\mbox{and}\quad
 \cov(n,n^\|)\, = \, 2\lepsd t\,.
\eeao

\proof
We calculate the expectation, the result for the covariance is then obvious.
By independence of the Poisson processes of joint and single jumps,
\beao
\Bbb E\left[n n^\|\right] &=& \Bbb E\left[(2n^\|+n^\bot_1+n^\bot_2) n^\|\right]\\
&=& 2\big(\var(n^\|) + (\bbe [n^\|])^2\big)+\Bbb E[n^\bot_1+n^\bot_2] \Bbb E[n^\|]  \\
&=&  2(\lepsd t + (\lepsd t)^2)+(\la_1^{(\eps)\bot}+\la_2^{(\eps)\bot})\lepsd t^2 \\
&=& 2\lepsd t (1+\leps t).
\eeao
\halmos
 \ele

 \ble\label{Mlemma}
Assume a bivariate $\al$-stable Clayton subordinator with common marginal parameters $(\log c,\al)$ and dependence parameter $\theta=\al\delta$.
 Assume also the observation scheme given in Section~\ref{s2}.
Define
\beam\label{T}
T(x,y):= (\log x+\log y)+\frac{\al}{\thet^2}\log(x^\thet+y^\thet)-(2+\frac{\al}{\thet})
 \frac{\p}{\p\thet}\log(x^\thet+y^\thet)\,.
 \eeam
Then {the matrix
 $M$ introduced in \eqref{mgoda} is given by }
 \beam\label{M}
 {M}=
 \left(\barr{ccc}
 1 &- \log\eps & 2 d \frac{\al\log2}{\thet^2}\\
  -\log\eps & \frac{1}{\al^2}+ (\log\eps)^2 & -d \Big(2\frac{\al\log 2}{\theta^2}\log\eps + m\Big)\\
2 d \frac{\al\log2}{\thet^2} &
-d \Big(2\frac{\al\log 2}{\theta^2}\log\eps + m\Big) & d b
 \earr
 \right),
 \eeam
 where $b$ is given by \eqref{b} and
\beam\label{m}
m=  2 \, \Cov\Big(\log\frac{X_1}{\eps}, T\Big(\frac{X_1}{\eps},\frac{Y_1}{\eps}\Big)\Big).
\eeam
Moreover, this covariance is independent of $\eps$.
\halmos
\ele

\subsection{Consistency and asymptotic normality of the two-step estimators}\label{s52}

Assume that the log-likelihood $l_{12}^{(\eps)}(\log c,\al)$  in
\eqref{like1} is used  for estimating  the marginal parameters
$\log c$ and $\al$ in the first step and the log-likelihood
$l^{(\eps)}(\log c,\al,\thet)$ in \eqref{like2} for estimating the
dependence parameter $\thet$ in the second step; i.e. we work with $ \bJ^{(\eps)}(\mathbf{X,Y};\eta)$ as given in Remark~\ref{rem2}.
As before we denote the resulting estimates by $\wt\ga = (\log\wt c,\wt\al)$
and $\wt\eta=(\log\wt c,\wt\al,\wt\theta)$.
Assume further that $\ga_0=(\log c_0,\al_0)$ and
$\eta_0=(\log c_0,\al_0,\thet_0)$ are the true parameter vectors.

Taylor expansions of each of the score functions in~\eqref{score1}
and~\eqref{score2} separately yield
 \beam\label{taylor}
 \barr{l}
 \frac{\p l_{12}^{(\eps)}(\log c,\al)}{\p\log c}\Big
 |_{\ga=\wt\ga}= \frac{\p l_{12}^{(\eps)}(\ga)}{\p\log c}\Big
 |_{\ga=\ga_0}+(\log \wt c-\log c_0)\,\frac{\p^2 l_{12}^{(\eps)}(\ga)}{\p{(\log c)}^2}\Big
 |_{\ga=\ga_{***}}+
 (\wt\al-\al_0)\,\frac{\p^2 l_{12}^{(\eps)}(\ga)}{\p\al\p\log c}\Big
 |_{\ga=\ga_{***}}\\
 \\
 \frac{\p l_{12}^{(\eps)}(\log c,\al)}{\p\al}\Big
 |_{\ga=\wt\ga}= \frac{\p l_{12}^{(\eps)}(\ga)}{\p\al}\Big
 |_{\ga=\ga_0}+(\log \wt c-\log c_0)\,\frac{\p^2 l_{12}^{(\eps)}(\ga)}{\p\log c\p\al}\Big
 |_{\ga=\ga_{\ast\ast}}+
 (\wt\al-\al_0)\,\frac{\p^2 l_{12}^{(\eps)}(\ga)}{\p\al^2}\Big
 |_{\ga=\ga_{\ast\ast}}\\
 \\
 \frac{\p {l}^{(\eps)}(\log c,\al,\thet)}{\p\thet}\Big
 |_{\eta=\wt\eta}=
 \frac{\p {l}^{(\eps)}(\eta)}{\p\thet}
 \Big |_{\eta=\eta_0}
 +(\log\wt c-\log c_0)\frac{\p^2 l^{(\eps)}(\eta)}{\p\log c\ \p\thet}\Big
 |_{\eta=\eta_{*}}+(\wt\al-\al_0)\frac{\p^2 l^{(\eps)}(\eta)}{\p\al\p\thet}\Big
 |_{\eta=\eta_{*}}\\
 \\
 \qquad\qquad\qquad\qquad+(\wt\thet-\thet_{0})\,\frac{\p^2 l^{(\eps)}(\eta)}{\p\thet^2}\Big
 |_{\eta=\eta_{\ast}}
 \earr
 \eeam
where $\ga_{***}$ and $\ga_{\ast\ast}$ are between
 $\wt\ga=(\log\wt c,\wt\al)$ and $\ga_0=(\log c_0,\al_0)$, and $\eta_{*}$ is between $\wt\eta$ and $\eta_0$, componentwise.

 Since the left hand sides of the equations in \eqref{taylor} are zero, so are the
 equations on the right hand side.
 Recall from Lemma~\ref{Heps} the matrix $H^{(\eps)}=H^{(\eps)}(\eta)$ of the second-order derivatives of
 the log-likelihoods and denote by
 $H_*^{(\eps)}$ the matrix $H^{(\eps)}$ at $\ga_{***}$, $\ga_{\ast\ast}$ and $\eta_{*}$ row-wise.
 Recalling the vector of score functions $\mathbf{J}^{(\eps)}(\eta)$ from Remark~\ref{rem2}, we rewrite \eqref{taylor} as
 \beam\label{Taylor2}
 H_*^{(\eps)}\ (\wt\eta-\eta_0)=-\mathbf{J}^{(\eps)}(\eta)\Big|_{\eta=\eta_0}.
 \eeam
Rewrite the components of the vector $\mathbf{J}^{(\eps)}(\eta)$ of the score functions
 in  \eqref{score1} and \eqref{score2} as
 \beam\label{score12}
 \frac{\p l_{12}^{(\eps)}(\log c,\al)}{\p\log c}&=&2\la^{(\eps)}t \,
 \Big(\frac{\wh\la^{(\eps)}}{\la^{(\eps)}}-1\Big)\\
 \label{score122}
 \frac{\p l_{12}^{(\eps)}(\log c,\al)}{\p\al}&=& 2\la^{(\eps)}
 t\, \log\eps\,\Big(\frac{\wh\la^{(\eps)}}{\la^{(\eps)}}-1\Big)-\sum_{i=1}^{n}\Big(\log(\frac{Z_i}{\eps})-\frac{1}{\al}\Big)\\
 \label{score123}
 \frac{\p l^{(\eps)}(\log c,\al,\thet)}{\p \thet}&=& \sum_{i=1}^{n^\|} T_i + \frac{n^\|}{\al+\thet}-\lepsd
 t\,\frac{\al\log2}{\thet^2}\nonumber\\
 &=& \sum_{i=1}^{n^\|}
 \left(T_i-\mu_T\right) +\frac{\al\log2}{\thet^2}\lepsd t \Big(\frac{\wh\lambda^{(\eps)\|}}{\lepsd}-1\Big).
 \eeam

The next result shows the consistency of the estimator $\wt\eta$.

 \bpr\label{consistency}
 Assume the  bivariate $\al$-stable
Clayton subordinator with common marginal parameters $(\log
c,\al)$ and dependence parameter $\theta=\al\delta$. Assume also
the observation scheme as described in Section~\ref{s3}. Let
$b(\al,\theta)$ be defined as in \eqref{b} and assume that
$b(\al,\theta)\neq 0$. Then the two-step estimator $\wt\eta$ is
consistent; i.e., as $n^\|\to\infty$ (then also $n\to\infty$) for fixed $\eps>0$,
 $$\wt\eta\quad\stp\quad\eta.$$
\epr

 \proof We denote again by $\eta_0$ the true parameter vector.
 Now divide \eqref{Taylor2} by $n$ and obtain
 \beam\label{Taylor3}
 \frac{1}{n} H_*^{(\eps)}
 (\wt\eta-\eta_0)=-\frac{1}{n}\mathbf{J}^{(\eps)}(\eta)\Big|_{\eta=\eta_0}.
 \eeam
 From the equations in \eqref{score12}, \eqref{score122} and \eqref{score123} the vector on the right hand side of \eqref{Taylor3} is given by
 \beao
 \frac{1}{n}\mathbf{J}^{(\eps)}(\eta)\Big|_{\eta=\eta_0}
 = \frac{2\lambda^{(\eps)}t}{n}
 \left(\barr{c}
 \frac{\wh\la^{(\eps)}}{\la^{(\eps)}}-1\\ \log\eps\,\Big(\frac{\wh\la^{(\eps)}}{\la^{(\eps)}}-1\Big)-\frac1{2\la^{(\eps)}t}\sum_{i=1}^{n}\Big(\log(\frac{Z_i}{\eps})-\frac{1}{\al}\Big)\\
 \frac1{2\la^{(\eps)}t}\sum_{i=1}^{n^\|} \left(T_i-\mu_T\right)+\frac{\al\log2}{\thet^2}\frac{\lepsd}{2\leps}
 \Big(\frac{\wh\la^{\eps)\|}}{\lepsd}-1\Big)
 \earr\right)_{\eta=\eta_0}.
 \eeao
Now, by the  Marcinkiewicz-Zygmund SLLN (cf. e.g.~\cite{EKM}, Theorem~2.5.10)
$\wh\lambda^{(\eps)}=\frac{n}{2t}\stas\leps_0$ as $n\to\infty$.
Invoking the same argument for $\wh\la^{(\eps)\|}$ ensures that the second summand of the third component tends to 0 a.s. as $n^{\|}\to\infty$.
For the terms involving $Z_i$ and $T_i$ we apply the SLLN for random sums (cf. e.g. \cite{EKM}, Lemma~2.5.3) and obtain that the right hand side of \eqref{Taylor3} tends to the zero vector a.s. as $n^{\|}\to\infty$ (and $\nto$).

Next we show that the limit of $\frac1{n} H^{(\eps)}(\eta_0)$
exists and is  deterministic and independent of $t$ as
$n^\|\to\infty$. Note that by the SLLN for random sums
$A(\al_0,\theta_0)$ and $B(\al_0,\theta_0)$ converge a.s. to
$a(\al_0,\theta_0)$ and $b(\al_0,\theta_0)$ as defined in
\eqref{a} and \eqref{b}, respectively. Using also the fact that
$2\la^{(\eps)}_0 t/n\stas 1$, we obtain $\frac1{n}
H^{(\eps)}(\eta_0) \stas -D$, where $D$ is given in \eqref{DstableC}
 and is invertible by Remark~\ref{Dinv}.

Now consider $H_*^{(\eps)}$ and write
$\leps_{***}=\leps\big|_{\ga_{***}}$, $\leps_{**}=\leps\big|_{\ga_{**}}$,
$\leps_{*}=\leps\big|_{\eta_{*}}$, $\al_{**}=\al\big|_{\ga_{**}}$, $\al_{*}=\al\big|_{\eta_{*}}$
$A_{*}=A\big|_{\eta_{*}}$,  $B_{*}=B\big|_{\eta_{*}}$, and $d_{*}=d\big|_{\eta_{*}}$.
Then we calculate
$${\rm det}H_*^{(\eps)}=- 4 \frac{d_{*} B_{*}}{\al_{**}^2} \leps_{***} \leps_{*} t^2 n \neq 0$$
for $\eta$ close to $\eta_0$ by continuity.
Hence, $H_*^{(\eps)}$ is invertible.
From \eqref{Taylor3} we obtain
 \beam\label{Hinverse}
 \left(
 \frac{\log\wt c-\log c_0}{\log\eps},
 \wt\al-\al_0,
 \wt\thet-\thet_0
 \right)^\top
= - (\frac1{n} H_*^{(\eps)})^{-1} \times\frac1{{n}}\mathbf{J}^{(\eps)}(\eta)\Big|_{\eta=\eta_0},
 \eeam
 where
 \beao
\frac1{n} H_*^{(\eps)} =
 \left(\barr{ccc}
 \frac{\leps_{***}}{\wh\la^{(\eps)}} & 0 & 0\\
 0 & \frac{\leps_{**}}{\wh\la^{(\eps)}} & 0\\
 0 & 0 & \frac{\la_{*}^{(\eps)}}{\wh\la^{(\eps)}}
  \earr\right)
  \times
 \left(\barr{ccc}
 -\log\eps & \log\eps & 0\\
 (\log\eps)^2 & -\frac{\wh\la^{(\eps)}}{\al_{**}^2 \la_{**}^{(\eps)}}- (\log\eps)^2 & 0 \\
 -d_* \frac{\al_*\log2}{\thet_*^2}\log\eps & d_*  \big( \frac{\al_*\log2}{\thet_*^2}\log\eps-A_* \big) & - d_*  B_*
  \earr\right).
  \eeao
  This implies that its inverse is given by
  \beao
(\frac1{n} H_*^{(\eps)})^{-1} =
  \left(\barr{ccc}
-\frac1{\log\eps}-\frac{\lepss}{\wh\la^{(\eps)}}\al_{**}^2\log\eps & -\frac{\lepss }{\wh\la^{(\eps)}}\al_{**}^2& 0\\
  -\frac{\lepss }{\wh\la^{(\eps)}}\al_{**}^2\log\eps & -\frac{\lepss}{\wh\la^{(\eps)}}\al_{**}^2& 0 \\
  \frac{A_*\al_{**}^2}{B_*} \frac{\lepss}{\wh\la^{(\eps)}}\log\eps+\frac{\al_{*}\log2}{\thet_*^2B_*}&
  \frac{A_*\al_{**}^2}{B_*} \frac{\lepss }{\wh\la^{(\eps)}}& -\frac1{d_* B_*}
 \earr\right)
 \times
  \left(\barr{ccc}
 \frac{\wh\la^{(\eps)}}{\leps_{***}} & 0 & 0\\
 0 & \frac{\wh\la^{(\eps)}}{\leps_{**}} & 0\\
 0 & 0 & \frac{\wh\la^{(\eps)}}{\leps_{*}}
  \earr\right).
 \eeao
All matrix elements are finite random variables and they
remain bounded in probability for $n^{\|}\to\infty$, since
$\wh\la^{(\eps)}\stas\la_0^{(\eps)}$ and all starred values are between the estimates
 and the true parameter values. Hence we
conclude that $\wt\eta\stp\eta_0$.
\eproof

We are now ready to formulate the main result of our paper.

\bthe\label{asnor}
Assume a bivariate $\al$-stable Clayton
subordinator with common marginal parameters $(\log c,\al)$ and
dependence parameter $\theta=\al\delta$. Assume also the
observation scheme as described in Section~\ref{s3}.
Let $a(\al,\theta)$ and $b(\al,\theta)$ be defined as in \eqref{a} and \eqref{b}, respectively,
and let $m$ be as in \eqref{m}.
If $b(\al,\theta)\neq 0$, then as $\eps\to0$,
 \beam\label{AN}
 \sqrt{2c\eps^{-\al}t}\left(\barr{c}
 \frac{\log\wt c-\log c}{\log\eps}\\
 \wt\al-\al\\
 \wt\thet-\thet
 \earr\right)
  \std N_3\left(\mathbf{0},V
  \right),
  \eeam
where
 \beam
 V = \left(\barr{ccc}
 \al^2 & \al^2 & -\frac{\al^2(a+m)}{b}\\
 \al^2 & \al^2 & -\frac{\al^2(a+m)}{b}\\
 -\frac{\al^2(a+m)}{b} & -\frac{\al^2(a+m)}{b} &
 \frac1{bd}-\frac{3\al^2(\log2)^2}{b^2\thet^4}+\frac{a\al^2(a+2m)}{b^2}
 \earr\right).
 \eeam
 \ethe

\proof
We start with the left hand side of equation in \eqref{Taylor3}.
 Multiplying both sides of \eqref{Taylor3} by ${\sqrt{n}}$
 yields as in the proof of Proposition~\ref{consistency} for $n^{\|}$ (hence $n$) sufficiently large by consistency that $H_*^{(\eps)}$ is invertible and by \eqref{Hinverse}
 \beam
 \sqrt{n}
 \left(
 \frac{\log\wt c-\log c_0}{\log\eps},
 \wt\al-\al_0,
 \wt\thet-\thet_0
 \right)^\top
&=& -(\frac1{n}H_*^{(\eps)})^{-1}
 \times
 \frac1{\sqrt{n}}\mathbf{J}^{(\eps)}(\eta)\Big|_{\eta=\eta_0}\nonumber\\
  \eeam
 The vector  $\frac1{\sqrt{n}}\mathbf{J}^{(\eps)}(\eta)$ is asymptotically normal with mean zero and covariance matrix
 $M={\frac1{2\leps t}}\Bbb E\big[\frac{\p l^{(\eps)}(\eta)}{\p\eta}\frac{\p l^{(\eps)}(\eta)}{\p\eta}^T\big]$
 calculated in Lemma \ref{Mlemma}.
 Since $\ga^{***}$ and $\ga^{**}$ are between $\wt\ga$ and $\ga_0$ and $\eta^{*}$
 between $\wt\eta$ and $\eta_0$, from the consistency of the
 estimators in Proposition~\ref{consistency}, $(H_*^{(\eps)})^{-1}$ converges a.s. by the SLLN and the consistency of the parameters to $D^{-1}$ as calculated in Remark~\ref{Dinv}.
 Consequently,
$\sqrt{n}\left(\frac{\log\wt c-\log c_0}{\log\eps}, \wt\al-\al_0, \wt\thet-\thet_0 \right)^\top$  converges as $n\to\infty$ in distribution to a normal vector with asymptotic covariance matrix
 \beao
 G^{-1} &=& D^{-1} M (D^{-1})^\top \Big |_{\eta=\eta_0}\\
& = &
 \left(\barr{ccc}
 \frac1{(\log\eps)^2}+\al_0^2 & \al_0^2&-\frac{(a_0+m_0)\al_0^2}{b_0}+\frac{\al_0\log2}{b_0\thet_0^2\log\eps}\\
  \al_0^2 & \al_0^2&  -\frac{(a_0+m_0)\al_0^2}{b_0}\\
 -\frac{(a_0+m_0)\al_0^2}{b_0}+\frac{\al_0\log2}{b_0\thet_0^2\log\eps}&-\frac{(a_0+m_0)\al_0^2}{b_0}
 & \frac1{b_0d_0}-\frac{3\al_0^2(\log2)^2}{b_0^2\thet_0^4}+\frac{a_0\al_0^2(a_0+2m_0)}{b_0^2}
 \earr\right)
 \eeao
with $G$ as in \eqref{godambe}.

Then by the SLLN we know that
 $\frac{\wh\la^{((\eps)}}{\leps}=\frac{n}{2\leps t}=\frac{n}{2c\eps^{-\al}t}\,\stas\,1$ as $n\to\infty$
 (either $\eps\downarrow0$ or $t\to\infty$), hence the rate $\sqrt{n}$ can be replaced by $\sqrt{2ct\eps^{-\al}}$.
Finally, $G^{-1}\to V$ as $\eps\to0$ and this completes the
proof.
 \halmos

\brem
(i) \, Note that for $t\to\infty$ and fixed $\eps>0$ the asymptotic covariance matrix in
\eqref{AN} is given by $G^{-1}$.\\[2mm]
(ii) \, The above theorem implies that the  normal limit vector has
representation
\beao (N_1,N_1,N_2)^\top, \eeao
where $N_1$ has
variance $\al^2$, $N_2$ has variance
 $\frac1{bd}+\frac{\al^2}{b^2}\Big(-\frac{3(\log2)^2}{\thet^4}+a(a+2m)\Big)$,
and the correlation between $N_1$ and $N_2$ is given by
$$\corr
(N_1,N_2)=-\frac{a+m}{\sqrt{\frac{1}{\al^2 d}-\frac{3(\log2)^2}{b\thet^4}+\frac{a}{b} (a+2 m) }}.$$
\erem

\bexam\label{ex-Godambe}
 [Asymptotic covariance matrix for a
bivariate $\al$-stable Clayton subordinator] \\
Let $\bS=(\bS(t))_{t\ge0}$ be a bivariate $\al$-stable
subordinator with a Clayton \lc\ as introduced in
Example~\ref{bss}. Assume further its parameters $c_1=c_2=c,\,
\al_1=\al_2=\al$ and $\thet=\al\de$ are estimated by a two-step
method as in Section~\ref{s42}.
The asymptotic covariance matrix as $\eps\to 0$ for the model with parameter values $c=1$, $\al=0.5$,
$\thet=1$ can be computed numerically similar to the calculation at the end of Section~5
in \cite{EsK2}. Note that it involves the numerical integration of certain integrals.
We find the asymptotic covariance matrix of
$\wt\eta=(\log\wt c,\wt\al,\wt\theta)$ as
$$V=\left[\barr{ccc}
  0.25 &   0.25 &   0.1042\\
    0.25  &  0.25 &  0.1042\\
  0.1042 &  0.1042  &  2.6273
 \earr\right].$$
 Alternatively, this matrix can also be estimated replacing the numerical integration by a Monte Carlo simulation.
For this the expectations in \eqref{a} and \eqref{b} and the
covariance $m$ from \eqref{m} are empirically estimated by
generating bivariate observations from \eqref{jointden1} with
parameter values mentioned above. Based on $10^6$ bivariate
observations, this yields the same asymptotic covariance matrix as
above (4 leading decimals coindice).

From this, we calculate $\corr(N_1,N_2)=0.1286$.
\eexam

\brem
For the bivariate $\al$-stable
Clayton \lp\ with equal marginal processes we have been able to calculate the Godambe information matrix analytically. However, for most
models this is too complicated. It requires
derivatives of first and second order, the integration of some
compound functions and the inverses and multiplications of
possibly high dimensional matrices.
As an alternative, a jackknife resampling method has been suggested
and can also be applied in this context for arbitrary \lp s;
cf. Joe~\cite{Joe:MultiModelsDepConcepts}, Section 10.1, and references given there.
\erem

\section{Maximum likelihood estimation of the full model}\label{s7}

We compare the two-step procedure presented in Section~\ref{s3}
with two alternatives. Firstly, we consider the estimation method
presented in \cite{EsK2} based on only common jumps. Secondly, we
also compare this method with the full likelihood, based on single
and common jumps. For this reason we present here the likelihood
function of the full model. The observation scheme is as explained
in Section~\ref{s3}, where $n^{\|}+n_1^{\bot}+n_2^{\bot}$ is the
number of observations.

From \eqref{jointLevy} the L\'evy densities of $\Pi_1^{(\eps)\bot}$, $\Pi_2^{(\eps)\bot}$ and $\Pi^{(\eps)\|}$ are given by
\beao
\nu_1^{\bot}(x) &=& c\al x^{-\al-1}\left[1-\left(1+({ x}/{\eps})^{-\al\de}\right)^{-1/\de-1}\right]\,,\quad x>\eps\\
\nu_2^{\bot}(y) &=& c\al y^{-\al-1} \left[1-\left(1+({y}/{\eps})^{-\al\de}\right)^{-1/\de-1}\right]\,,\quad y>\eps\,,\\
\nu^{\|}(x,y) &=& c \al^2 (1+\delta)  (xy)^{\al\delta-1}
\Big(x^{\al\delta}+y^{\al\delta}\Big)^{-1/\delta-2}\,. \eeao As
intensities we obtain from \eqref{jointint} $\la^{(\eps)\|}= c
2^{-1/\delta}\eps^{-\al}$. Moreover, the marginal intensities are
$\la^{(\eps)}_1=\la^{(\eps)}_2=c\eps^{-\al}$, so that
{$\la_1^{(\eps)\bot}=\la_2^{(\eps)\bot}=
c\eps^{-\al}(1-2^{-1/\delta})$.
This implies the intensity of the bivariate CPP
$\rho^{(\eps)} :=\la^{(\eps)\|}+\la_1^{(\eps)\bot}+\la_2^{(\eps)\bot}=
c\eps^{-\al}(2-2^{-1/\delta})$.}

For simplicity we reparameterize the model again as in
Section~\ref{s42} by setting $\al\delta=\theta$ and take {$\log
c$} instead of $c$ as second marginal parameter. Now we recall
Th.~4.1 of \cite{EsK1} for a bivariate CPP and obtain the
likelihood function; here $(x_i,y_i)_{i=1,\ldots,n^\|}$ denote the
common jumps in both components and $\wt x_i$ for
$i=1,\ldots,n_1^\bot$ and $\wt y_i$ for $i=1,\ldots,n_2^\bot$
denote the single jumps.
{The likelihood function of the bivariate CPP is then given by}
 \beam \label{lik-all}
 L^{(\eps)}(\log c,\al,\theta) &=&
\Big(e^{-\rho^{(\eps)} t}\prod_{i=1}^{n^{\|}}\nu^\|(x_i,y_i)\Big) \times
\Big(e^{-\la_1^{(\eps)\bot}
t}\prod_{i=1}^{n^{\bot}_1}\nu_1^{\bot}(\wt x_i)\Big) \times
\Big(e^{-\la_2^{(\eps)\bot}t} \prod_{i=1}^{n^{\bot}_2}\nu(\wt y_i)\Big)\nonumber\\
 &=&
{e^{-ct \eps^{-\al}(2-2^{-\al/\thet})}
(\al+\thet)^{n^\|}(\al c)^{n^\|+n_1^\bot+n_2^\bot}}
\prod_{i=1}^{n^\|}\left[(x_iy_i)^{\thet-1}(x_i^\thet+y_i^\thet)^{-\al/\thet-2}\right]\nonumber\\
&&
\times
\prod_{i=1}^{n_1^\bot}\left[\wt x_i^{-\al-1}\left(1-\left(1+(\wt x_i/\eps)^{-\thet}\right)^{-\al/\thet-1}\right)\right]\\
&&\times \prod_{i=1}^{n_2^\bot}\left[{\wt
y_i}^{-\al-1}\left(1-\left(1+(\wt
y_i/\eps)^{-\thet}\right)^{-\al/\thet-1}\right)\right].\nonumber
\eeam The log-likelihood is given by
 \beao
l^{(\eps)}(\log c,\al,\thet) &=&
{-ct\eps^{-\al} (2-2^{-\al/\thet})+ n^\| \log(\al+\thet)+(n^\|+n_1^\bot+n_2^\bot)(\log\al+\log c)}\\
&&+(\thet-1)
 \sum_{i=1}^{n^\|} (\log x_i+\log y_i)
 -(2+\frac{\al}{\theta})\sum_{i=1}^{n^\|}
\log(x_i^{\theta}+y_i^{\theta})\\
&&-(\al+1)\sum_{i=1}^{n_1^\bot}\log\wt
x_i+\sum_{i=1}^{n_1^\bot}\log\Big[1-\left(1+(\wt
x_i/\eps)^{-\thet}\right)^{-\al/\thet-1}\Big]\\
&&-(\al+1)\sum_{i=1}^{n_2^\bot}\log\wt
y_i+\sum_{i=1}^{n_2^\bot}\log\Big[1-\left(1+(\wt
y_i/\eps)^{-\thet}\right)^{-\al/\thet-1}\Big] \,.
\eeao
For the score functions we obtain
\beao
{\frac{\partial l^{(\eps)}}{\partial \log c}} &=& -ct\eps^{-\al} (2-2^{-\al/\thet})+\frac{n^\|+n_1^\bot+n_2^\bot}{c}\\
\frac{\partial l^{(\eps)}}{\partial\al } &=&
{ct\eps^{-\al}\Big(2\log\eps-2^{-\al/\thet}\log\eps-\frac{2^{-\al/\thet}\log2}{\thet}\Big)
+\frac{n^\|}{\al+\theta}
+\frac{n^\|+n_1^\bot+n_2^\bot}{\al}}\\
&&-\frac{1}{\thet}\sum_{i=1}^{n^\|}\log(x_i^{\theta}+y_i^{\theta})
 -\sum_{i=1}^{n_1^\bot}\log\wt x_i+
\sum_{i=1}^{n_1^\bot}\frac{\p}{\p\al}
\log\left[1-\left(1+(\wt x_i/\eps)^{-\thet}\right)^{-\al/\thet-1}\right]\\
&&- \sum_{i=1}^{n_2^\bot}\log\wt
y_i+\sum_{i=1}^{n_2^\bot}\frac{\p}{\p\al}\log\left[1-\left(1+(\wt y_i/\eps)^{-\thet}\right)^{-\al/\thet-1}\right]
 \eeao
 \beao
 \frac{\partial
l^{(\eps)}}{\partial \theta}
&=&{\frac{ct\al\eps^{-\al}2^{-\al/\thet}\log2}{\thet^2}
}+\frac{n^\|}{\al+\theta}
 +\sum_{i=1}^{n^\|}(\log x_i+\log y_i)
+\frac{\al}{\thet^2}\sum_{i=1}^{n^\|}\log(x_i^\thet+y_i^\thet)\\
 &&
 - (2+\frac{\al}{\theta})\sum_{i=1}^{n^{\|}}
\frac{\partial}{\partial \theta}\log(x_i^{\theta}+y_i^{\theta})+
\sum_{i=1}^{n_1^\bot}\frac{\p}{\p\thet}\log\left[1-\left(1+(\wt
x_i/\eps)^{-\thet}\right)^{-\al/\thet-1}\right]\\
&&+\sum_{i=1}^{n_2^\bot}\frac{\p}{\p\thet}\log\left[1-\left(1+(\wt
y_i/\eps)^{-\thet}\right)^{-\al/\thet-1}\right] \,.
\eeao
The three parameters are obtained by numerical optimization.

It is possible to prove joint asymptotic normality of $(\log c,\al,\theta)$ similar to our calculations in Esmaeili and Kl\"uppelberg~\cite{EsK2}
and in Section~\ref{s4} of the present paper.
However, for the observation scheme of the present paper this is even more complicated than in~\cite{EsK2}.
We refrain from this tedious analytic exercise and, instead, present the results of a simulation study in the next section,
where we compare all three methods presented.

\section{Comparison of estimation procedures}\label{s8}

\begin{table}[t]
\begin{center}
\begin{tabular*}{146.2mm}{@{}|c|c|c|c|c|c |@{}}
 \cline{1-6}
  Method of estimation &Truncation point& & $c=1$ & $\al=0.5$ & $\de=2$  \\
 \cline{1-6}
     &  & Mean &   1.0678    &    0.5289    &     2.1489 \\
 \cline{3-6}
  &$\eps=0.001$&$\sqrt{MSE}$ &   0.6344   & 0.1206   & 0.9511 \\
 \cline{3-6}
   MLE&&$MRB$& 0.0517 &   0.0525 &   0.0842\\
 \cline{2-6}
     (only bivariate jumps)& &Mean &  1.0460   &      0.5020   &    2.0301 \\
 \cline{3-6}
 as in \cite{EsK2}
  &$\eps=0.00001$&$\sqrt{MSE}$ &  0.3677  &  0.0349  &  0.2488\\
 \cline{3-6}
  &&$MRB$&  0.0413  &  0.0044  &  0.0144\\
 \cline{1-6}
 &&Mean &   1.0177 &        0.5216    &    2.0129  \\
 \cline{3-6}
 &$\eps=0.001$&$\sqrt{MSE}$ &  0.5248  &  0.0777 &   0.4337 \\
 \cline{3-6}
 MLE&&$MRB$& 0.0072  &  0.0423  &  0.0119\\
 \cline{2-6}
 (full model)& &Mean &   1.0175&    0.5021&    2.0091  \\
 \cline{3-6}
 as in Section~\ref{s4}
 &$\eps=0.00001$&$\sqrt{MSE}$ &  0.2808  &  0.0239 &   0.1253 \\
 \cline{3-6}
 &&$MRB$&   0.0142  &  0.0045  &  0.0042\\
 \cline{1-6}
 &&Mean & 1.0453    &  0.5231  &      2.0762 \\
 \cline{3-6}
 IFM
  &$\eps=0.001$ &$\sqrt{MSE}$ &  0.5535  &  0.0859  &  0.6764 \\
\cline{3-6}
  (two-step method) &&$MRB$&  0.0264  &  0.0471  &  0.0379\\
  \cline{2-6}
  as in Section~\ref{s42}
   & &Mean &   1.0301 &   0.5021  &   2.0149  \\
 \cline{3-6}
  &$\eps=0.00001$&$\sqrt{MSE}$ &  0.3003  &  0.0257  &  0.1696\\
\cline{3-6}
 &&$MRB$&     0.0249 &   0.0048  &  0.0065\\
 \cline{1-6}
 \end {tabular*}\\[0.5ex]
 \end{center}
 \caption{\label{table1}
Comparison of estimates for a bivariate $\frac{1}{2}$-stable
Clayton process with common marginal parameters.
We simulated 100 sample paths and estimated all parameters 100 times.
Each of the 100 estimates was based on one sample path, on which all three methods were performed.
From each sample path we truncated the small jumps based on the two truncation points ($\eps=0.001$ and $\eps=0.00001$), respectively.
Each sample path of the process was
simulated as a continuous time realization of a CPP in one unit of time, $0\le
 t<1$, for $\tau=1000$, equivalent to truncation of the small jumps at the cut-off point $\xi=\ov\Pi^{\,\leftarrow}(\tau)=10^{-6}$.
 }
\end{table}

In this section we compare the quality of the MLEs
$\wh\eta=(\log\wh c,\wh\al,\wh\theta)$ of the full model of
Section~\ref{s7} with the estimates $\wt\eta=(\log\wt
c,\wt\al,\wt\theta)$ obtained by the two-step method in
Section~\ref{s5}. Moreover, we also include in our comparison
those estimates obtained from bivariate jumps larger than $\eps$
only as derived in {Th.~4.6 of \cite{EsK2}}. Since this last
method means to base the statistical analysis on less data, we
expect that this method is less efficient than the MLE based on
all available data. More precisely, for the first two parameters
$\log c$ and $\al$, the rate has simply changed from
{$\sqrt{c2^{-\al/\theta}\eps^{-\al}t}$} to {$\sqrt{c 2
\eps^{-\al}t}$}.

\subsubsection*{The simulation study}

We simulate sample paths of the bivariate $\al$-stable Clayton
subordinator with equal marginals and parameters given by  $c=1$
($\log c=0$), $ \al=1/2$ and $\de=2$ ($\thet=1$). We generate
sample paths of this process over a time span $[0,t]$, where we
choose $t=1$ for simplicity. Recall from our observation scheme
introduced in Section~\ref{s2} that we observe all jumps larger
than $\eps$ either in one component or in both. Obviously, we
cannot simulate a trajectory of a stable process, since we are
restricted to the simulation of a finite number of jumps. For
simulation purposes we choose a threshold $\xi$ (which should be
much smaller than $\eps$) and simulate jumps larger than $\xi$ in
one component, and arbitrary in the second component. To this end
we invoke Algorithm~6.15 in Cont and Tankov~\cite{ContTankov}.

The simulation of a bivariate $\al$-stable Clayton subordinator is
explained in detail in Example~6.18 of \cite{ContTankov}. The
algorithm starts by fixing a number $\tau$ determined by the
required precision. This number coincides with the jump intensity $\lx_1$, which fixes the average number of terms in the approximating CPP.
More details
can be found in \cite{EsK2}.

For the estimation we first consider $\eps=0.001$, i.e. a
relatively large truncation point. Not surprisingly, the MLEs based on the full model discussed in
Section~\ref{s7} are definitely better than the other estimates in
Table~\ref{table1}. We find it, however, remarkable that the
two-step method outperforms the MLE based on joint jumps only. The
reason for this is presumably that the MLE's based only on joint jumps use
only such data with \lm\ on $[\eps,\infty)^2$. The two-step method,
however, uses also data, which are only in one component larger
than $\eps$ in its first step. The marginal parameters are based on substantially more data.

\begin{figure}
\begin{center}
\begin{tabular}{ccc}
$\log c=0$& $\al=0.5$& $\theta =1$\\
\includegraphics[scale=0.30]{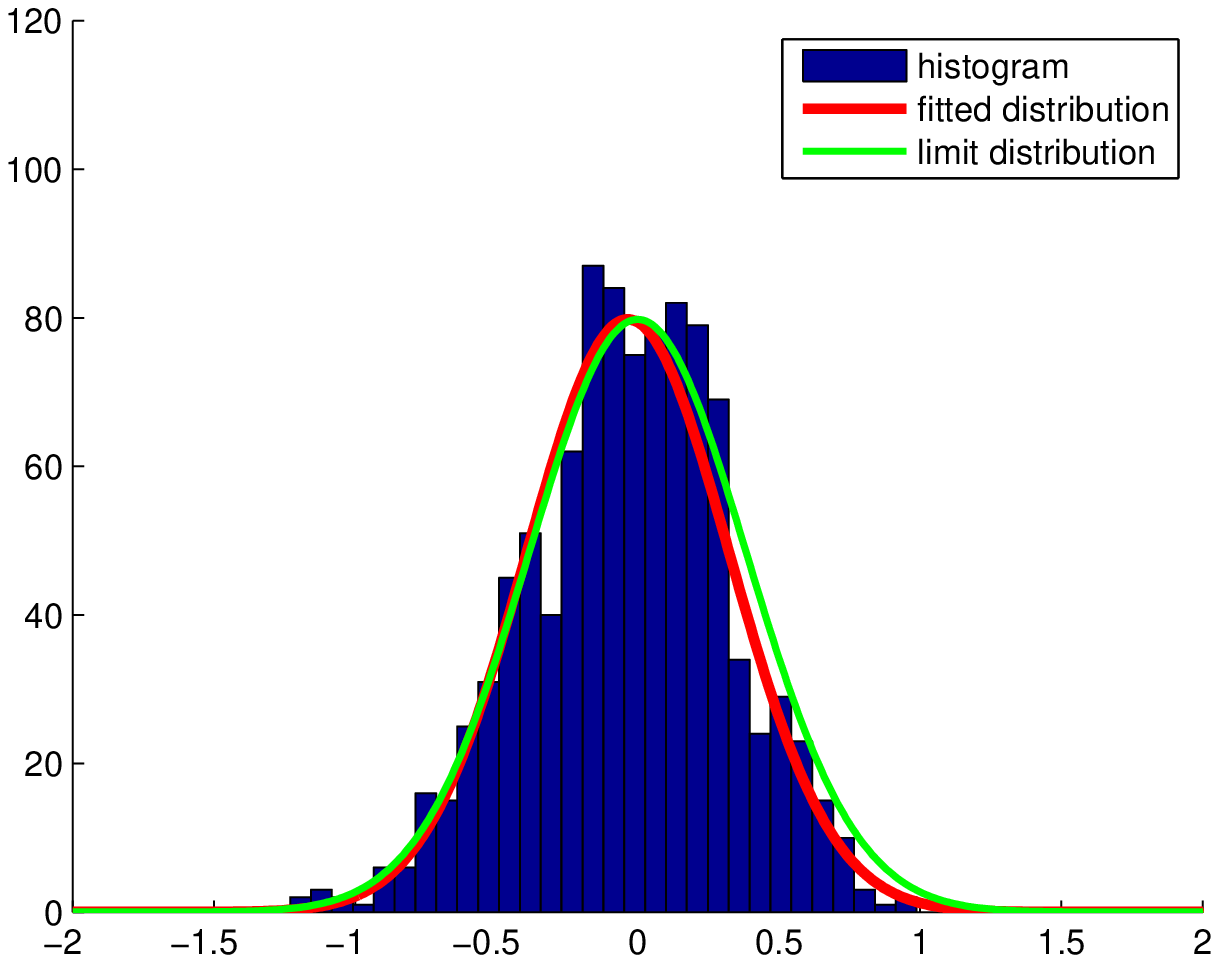} &
\includegraphics[scale=0.30]{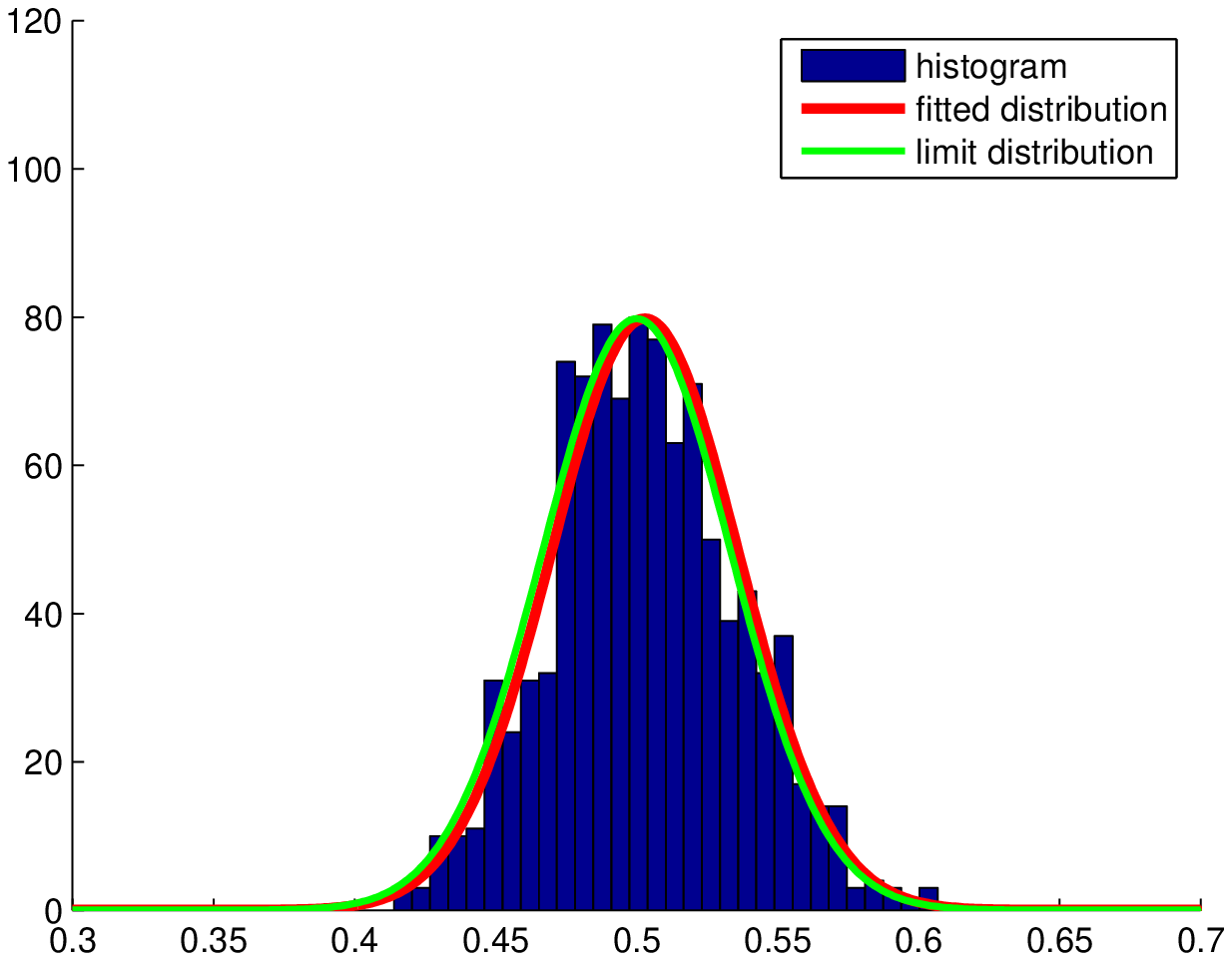}&
\includegraphics[scale=0.30]{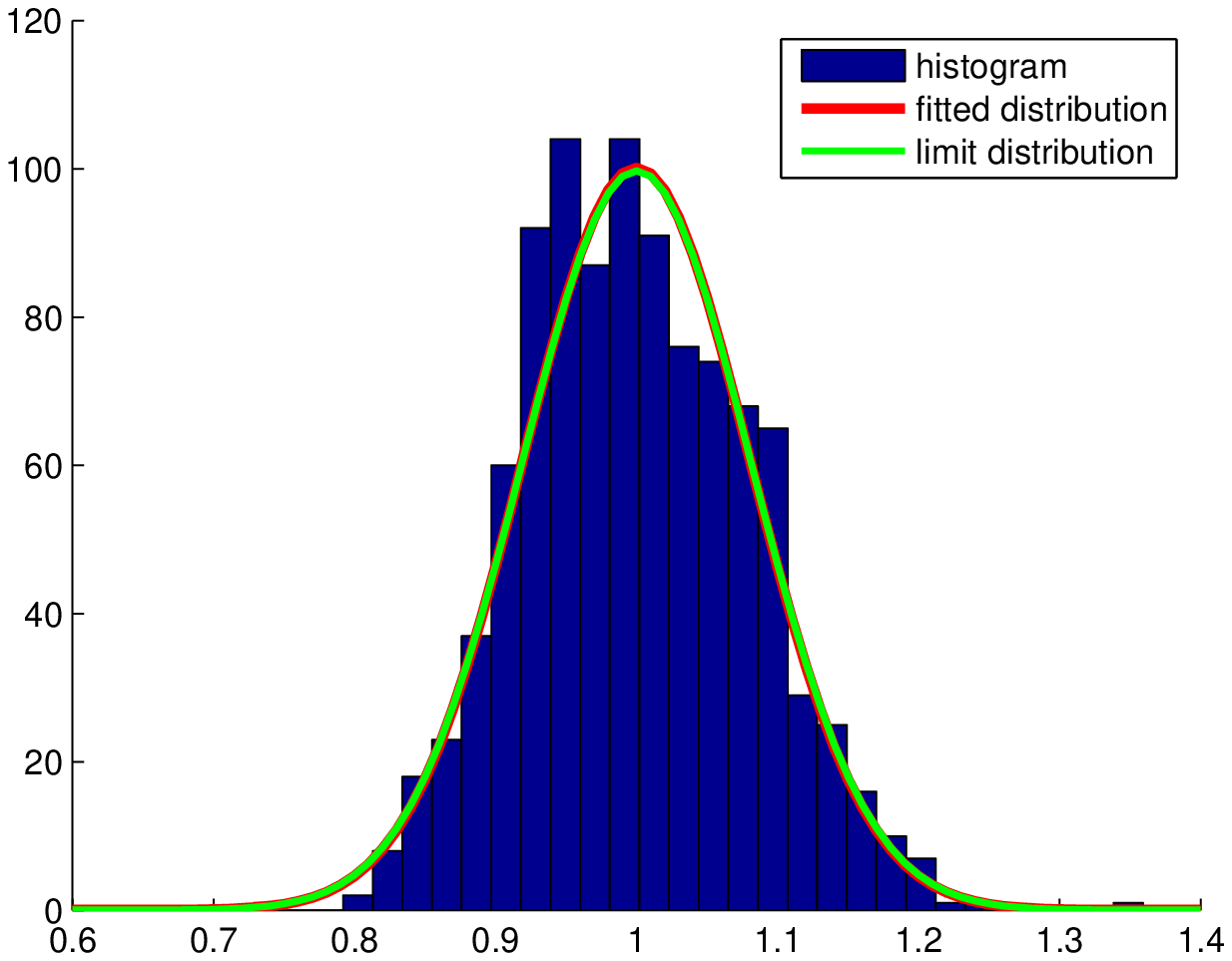}\\
\includegraphics[scale=0.30]{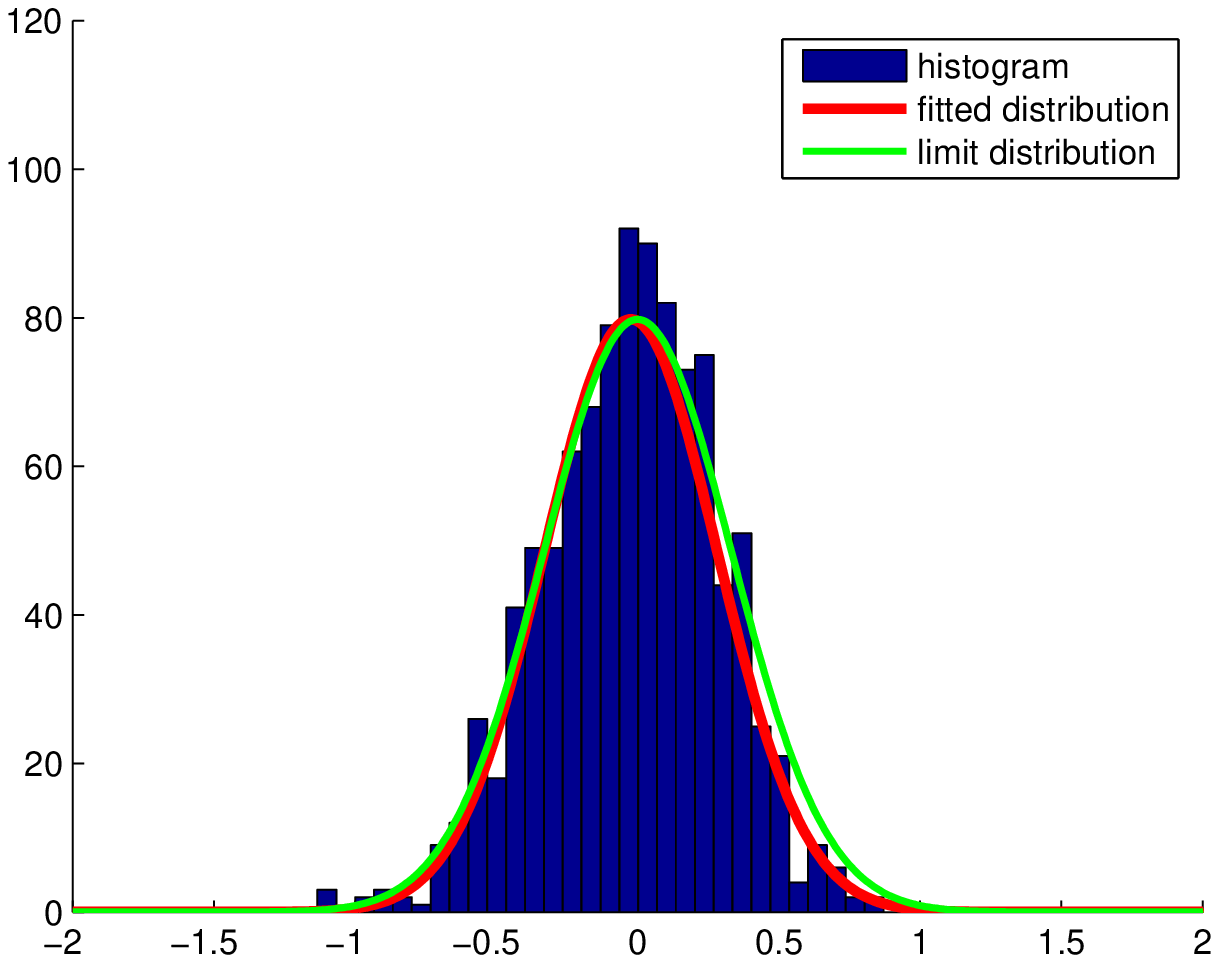}&
\includegraphics[scale=0.30]{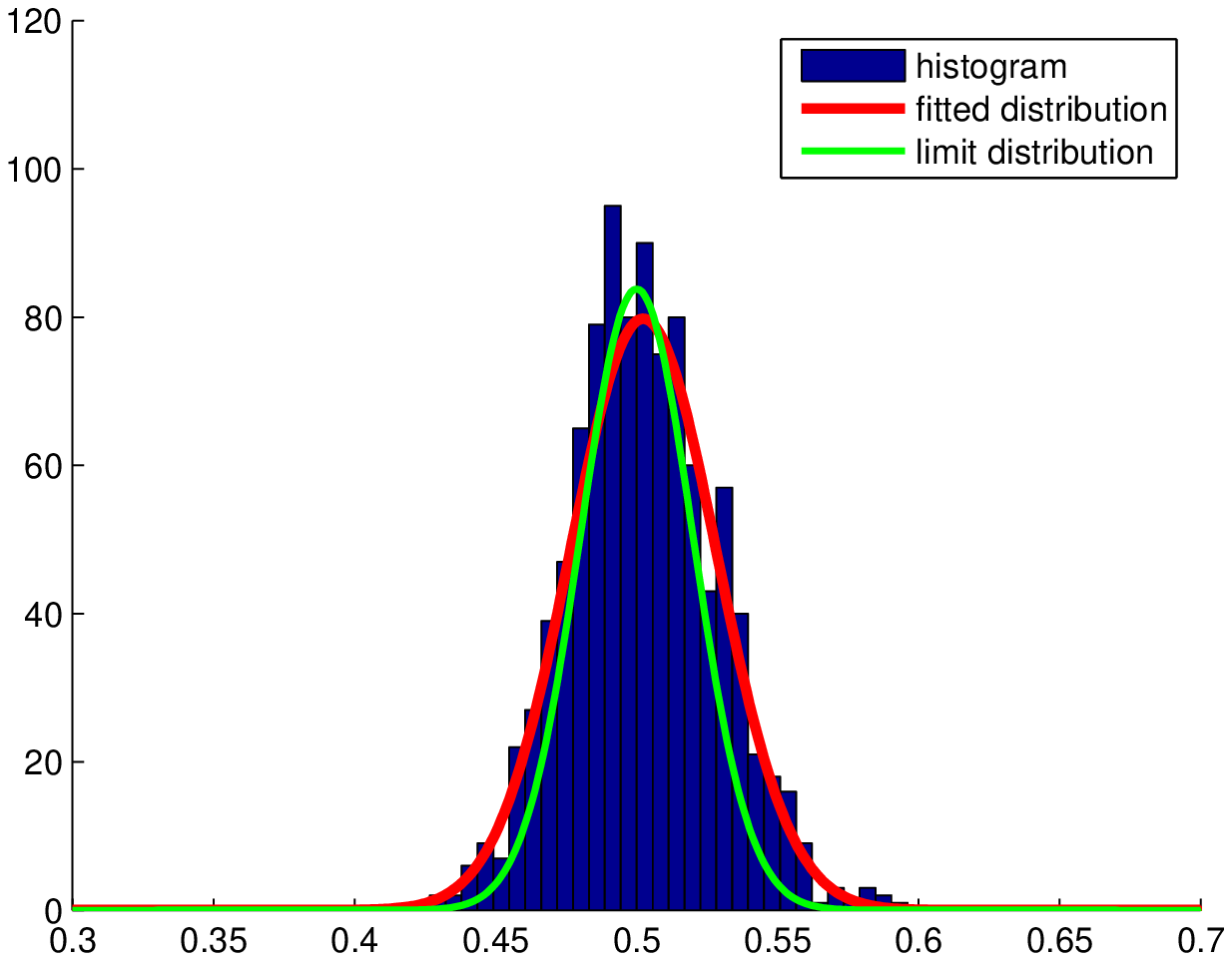}&
\includegraphics[scale=0.30]{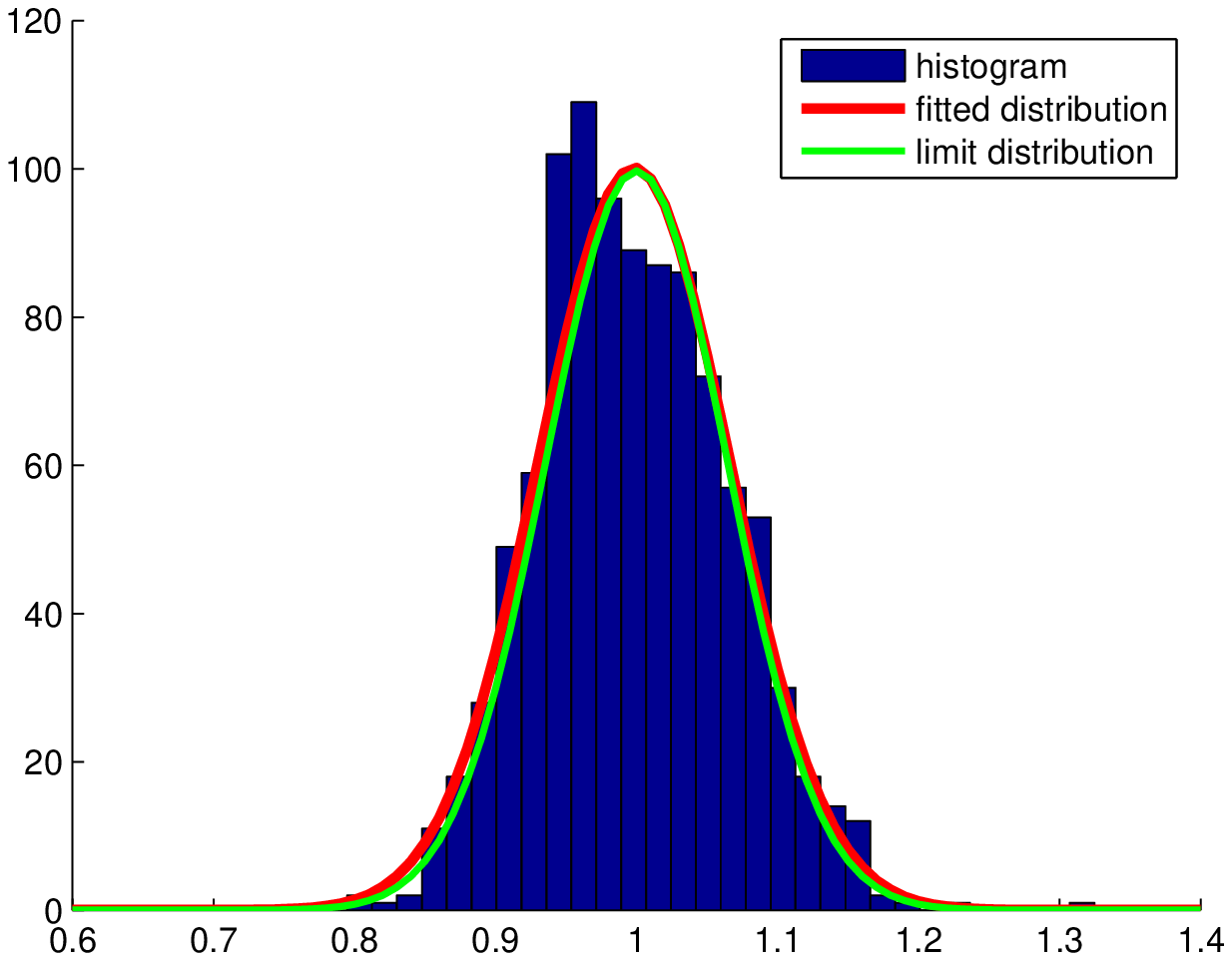}\\
\includegraphics[scale=0.30]{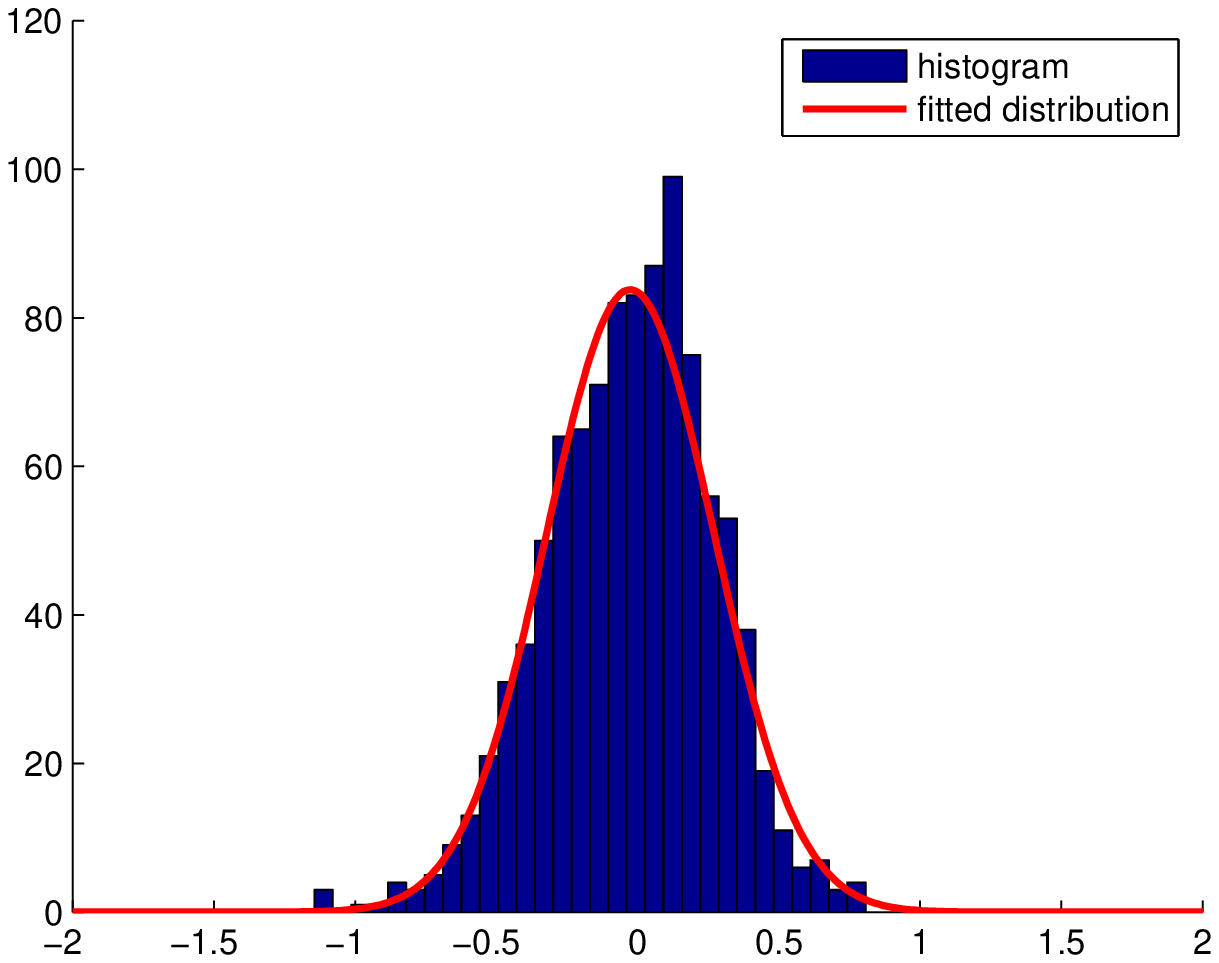}&
\includegraphics[scale=0.30]{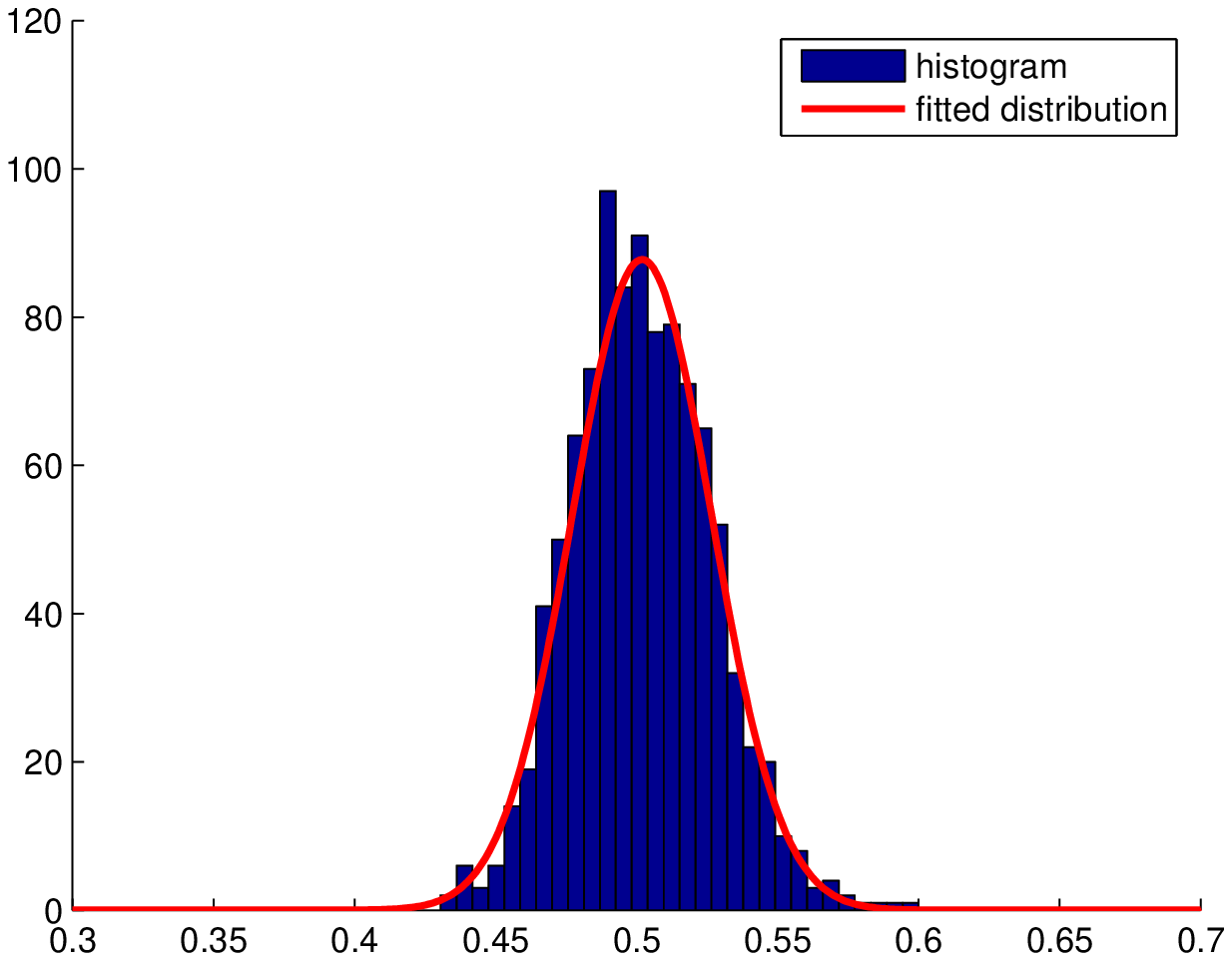}&
\includegraphics[scale=0.30]{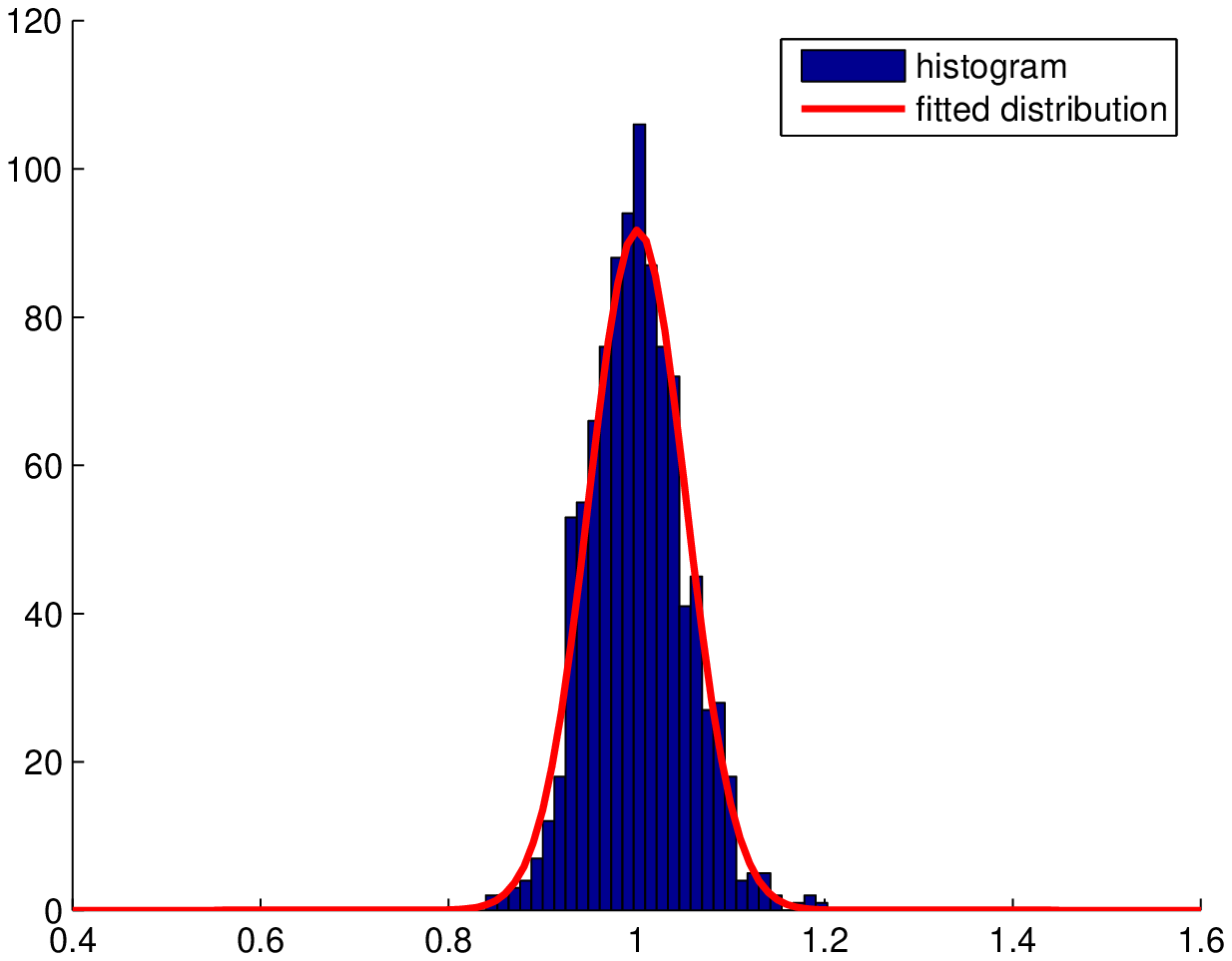}
\end{tabular}\\
\caption{\label{figure1}
 Histogram with statistically fitted normal density (red) and theoretical limit distribution (green) for 1000 parameter estimates of a bivariate Clayton stable \lp. The parameter values are $c=1$, $\al=0.5$ and $\delta=2$ and the jump-truncated point is $\eps=0.00001$.
The estimation procedures are MLEs based on joint jumps only (first row, limit distribution derived in Theorem~4.6 of \cite{EsK2}),
 the two-step method (second row, limit distribution derived in Theorem~\ref{asnor} above) and MLEs based on all jumps (third row, without theoretical limit law).}
\end{center}
\end{figure}

 When we consider also smaller jumps; i.e., if we choose $\eps=10^{-5}$, the estimates will be more
 precise with less variation and smaller bias. In Table~\ref{table1},
 the results in the lower part of each estimation method  show this fact.
 It can also be seen from this table that the MLEs from a full
 model have the least mean relative bias (MRB) and  mean
 square errors (MSE) as expected.
 In Figure~\ref{figure1} we visualize the situation for the this jump truncation point of $\eps=10^{-5}$ based on 1000 simulated sample paths. Again all three estimation methods are performed for each sample path.

\subsubsection*{Acknowledgement}
Both authors are grateful to Alexander Lindner for pointing out some inconsistencies in the proof or Proposition~\ref{consistency}.
His remarks lead also to an improvement of the presentation of our results.
C.K. gratefully acknowledges financial support from the Institute for Advanced Study of the Technische Universit\"at M\"unchen (TUM-IAS).

\bibliographystyle{plain}

\begin{thebibliography}{10}

\bibitem{BaB1}
I.V. Basawa and P.J. Brockwell.
\newblock Inference for gamma and stable processes.
\newblock {\em Biometrika}, 65(1):129--133, 1978.

\bibitem{BaB2}
I.V. Basawa and P.J. Brockwell.
\newblock A note on estimation for gamma and stable processes.
\newblock {\em Biometrika}, 67(1):234--236, 1980.

\bibitem{ContTankov}
R.~Cont and P.~Tankov.
\newblock {\em Financial Modelling with Jump Processes}.
\newblock Chapman \& Hall, Boca Raton, 2004.

\bibitem{EdeKlu08}
I.~Eder and C.~Kl\"uppelberg.
\newblock Pareto L\'evy measures and multivariate regular variation.
\newblock {\em Adv. in Appl. Probab.}, 44(1):117-138, 2012.

\bibitem{EKM}
P.~Embrechts, C.~Kl\"uppelberg and T.~Mikosch.
\newblock {\em Modelling Extremal Events for Insurance and Finance.}
\newblock Springer, Heidelberg, 1997.

\bibitem{EsK1}
H.~Esmaeili and C.~Kl\"uppelberg.
\newblock Parameter estimation of a bivariate compound poisson process.
\newblock {\em Insurance: Mathematics and Economics}, 47:224--233, 2010.

\bibitem{EsK2}
H.~Esmaeili and C.~Kl\"uppelberg.
\newblock Parameter estimation of a bivariate stable {L}\'evy process.
\newblock {\em J. Mult. Anal.}, 102(5):918--930, 2011.

\bibitem{God91}
V.P. Godambe.
\newblock {\em Estimating Functions.}
\newblock Oxford University Press, Oxford, 1991.

\bibitem{HJ}
R.~H\"opfner and J.~Jacod.
\newblock Some remarks on the joint estimation of the index and the scale
  parameter for stable processes.
\newblock In: P.~Mandl and M.~Huskova (Eds.) {\em Asymptotic Statistics.
  Proceedings of the Fifth Prague Symposium 1993}, pp. 273--284. Physica
  Verlag, Heidelberg, 1994.

\bibitem{Joe:MultiModelsDepConcepts}
H.~Joe.
\newblock {\em Multivariate Models and Dependence Concepts.}
\newblock Chapman \& Hall/CRC, London, 1997.

\bibitem{KalTan06}
J.~Kallsen and P.~Tankov.
\newblock Characterization of dependence of multidimensional {L\'evy} processes
  using {L\'evy} copulas.
\newblock {\em J. Mult. Anal.}, 97:1551--1572, 2006.

\bibitem{resnick:1987}
S.~I. Resnick.
\newblock {\em Extreme Values, Regular Variation, and Point Processes}.
\newblock Springer, New York, 1987.

\bibitem{Sato:1999}
K.~Sato.
\newblock {\em L\'evy Processes and Infinitely Divisible Distributions.}
\newblock Cambridge University Press, Cambridge, U.K., 1999.

\bibitem{Vaart}
Van der Vaart, A. W.
{\em Asymptotic Statistics.}
Cambridge University Press,
Cambridge, 2007.

\bibitem{Xu96}
J.J. Xu.
\newblock {\em Statistical Modelling and Inference for Multivariate and
  Longitudinal Discrete Response Data.}
\newblock Ph.D. Thesis, University of British Columbia, Department of
  Statistics, 1996.

\end{thebibliography}

\section{Appendix}\label{A}

{\em Proof of Lemma~\ref{Heps}. }
The score functions in \eqref{score1} have derivatives
\beam\label{secderstep1}
\frac{\p^2l_{12}^{(\eps)}(\log c,\al)}{\p(\log c)^2}&=&-2ct\eps^{-\al} = -2\leps t\nonumber\\
\frac{\p^2l_{12}^{(\eps)}(\log c,\al)}{\p\al\p\log c}&=&\frac{\p^2l_{12}^{(\eps)}(\log c,\al)}{\p\log c\ \p\al}=2ct\eps^{-\al}\log\eps=2\leps t \log\eps\\
\frac{\p^2l_{12}^{(\eps)}(\log c,\al)}{\p\al^2}&=&-\frac{n}{\al^2}-2ct\eps^{-\al}({\log \eps})^2
=-\frac{n}{\al^2}-2\leps t (\log \eps)^2.\nonumber
\eeam
This means that the upper left $2\times 2$-matrix of $H^{(\eps)}$ is the Fisher information matrix to the MLE of $(\log c,\alpha)$, calculated by Basawa and Brockwell~\cite{BaB2}, and also presented in Esmaeili and Kl\"uppelberg~\cite{EsK2}, Example~3.1 (up to a deterministic factor), since here all observations from both marginals are considered.
Since the score functions in \eqref{score1} are independent of
 the parameter $\thet$, the matrix $H^{(\eps)}$ has the structure as given in \eqref{DstableC}.
It remains to calculate the last row of $H^{(\eps)}$.
We calculate the derivatives of the score function in \eqref{score2} as follows: first, by Lemma~\ref{lepsdder},
\beao
 \frac{\p^2 l^{(\eps)}(\log c,\al,\thet)}{\p\log c\p\thet}&=&
 -\frac{\p^2\lepsd}{\p\log c \p \thet} t =-\lepsd t \frac{\al\log2}{\thet^2}.
\eeao
Furthermore,
\beam\label{secderstep2}
 \frac{\p^2 l^{(\eps)}(\log c,\al,\thet)}{\p\al\p\thet}&=&
 -\frac{\p^2\lepsd}{\p\al\p\thet}t
 -\frac{n^\|}{(\al+\thet)^2}
+\frac{1}{\thet^2}\sum_{i=1}^{n^\|}\log{(X_i^\theta+Y_i^\theta)}\nonumber\\
&& -\frac{1}{\thet}\sum_{i=1}^{n^\|}\frac{\p}{\p\thet}
 \log{(X_i^\thet+Y_i^\thet)}
 \\
\frac{\p^2 l^{(\eps)}(\log c,\al,\thet)}{\p\thet^2}&=&
-\frac{\p^2\lepsd}{\p\thet^2}t
-\frac{n^\|}{(\al+\thet)^2}
-\frac{2\al}{\thet^3}\sum_{i=1}^{n^\|}\log{(X_i^\thet+Y_i^\thet)}\nonumber\\
&&+\frac{{2\al}}{\thet^2}\sum_{i=1}^{n^\|}\frac{\p}{\p\thet}\log{(X_i^\thet+Y_i^\thet)}
-(2+\frac{\al}{\thet})\sum_{i=1}^{n^\|}\frac{\p^2}{\thet^2}\log{(X_i^\thet+Y_i^\thet)}.\nonumber
 \eeam
\eproof

{\em Proof of Lemma~\ref{D}. } Since $D=-\frac1{2\leps t}\bbe[H^{(\eps)}]$,
it remains to calculate the expectations:
\beao
 \bbe[A(\al,\theta)] &=&
  - \frac{\al(\log2)^2}{\theta^3}+\frac{\log 2}{\theta^2}+\frac1{(\al+\thet)^2} -\frac{1}{\thet^2}
\Bbb E\Big[\log(X_1^\theta+Y_1^\theta)\Big]\\
 &&  + \frac{1}{\thet}\Bbb E\Big[\frac{\p}{\p\thet}\log{(X_1^\thet+Y_1^\thet)}\Big]\\
&=& a(\al,\theta)
 \\
 \bbe[B(\al,\theta)]  &=&
\Big(\frac{\al\log 2}{\theta^2}\Big)^2 -\frac{2\al\log 2}{\theta^3} +\frac{1}{(\al+\thet)^2}
 + \frac{2\al}{\thet^3}\Bbb E\Big[\log{(X_1^\thet+Y_1^\thet)}\Big]\\
 &&  -\frac{2\al}{\thet^2}\Bbb E\Big[\frac{\p}{\p\thet}\log{(X_1^\thet+Y_1^\thet)}\Big]
+(2+\frac{\al}{\thet})\Bbb E\Big[\frac{\p^2}{\p\thet^2}\log{(X_1^\thet+Y_1^\thet)}\Big]\\
&=& b(\al,\theta).
\eeao
\halmos

\noindent
{\em Proof of Lemma~\ref{Mlemma}. }
Recall the definition of the $Z_i$ for $i=1,\ldots,n$ as in
Step 1 of Section~\ref{s42} and the fact that $\log
\frac{Z_1}{\eps},\ldots,\log \frac{Z_n}{\eps}$ are exponential random variables with expectation $\alpha^{-1}$, and that in this first step they are treated as independent. The entries of
the matrix $M=(m_{ij})_{1\le i,j\le 3}$ are calculated from the
score functions \eqref{score-func2} in Remark~\ref{rem2} as
follows:
 \beao
  m_{11}&=&{\frac1{2\leps t}}\Bbb E\Big[\Big(\frac{\p l_{12}^{(\eps)}}{\p\log
 c}\Big)^2\Big]={\frac1{2\leps t}}\Bbb E\Big[-\frac{\p^2 l_{12}^{(\eps)}}{\p(\log
 c)^2}\Big] =d_{11}\\
m_{12}&=&{\frac1{2\leps t}}\Bbb E\Big[\Big(\frac{\p
 l_{12}^{(\eps)}}{\p\al}\Big)\Big(\frac{\p
 l_{12}^{(\eps)}}{\p\log c}\Big)\Big] ={\frac1{2\leps t}}\Bbb E\Big[-\frac{\p^2
 l_{12}^{(\eps)}}{\p\al\p\log c}\Big]
 = d_{12} = d_{21} = m_{21}\\
m_{22}&=&{\frac1{2\leps t}}\Bbb E\Big[\Big(\frac{\p
l_{12}^{(\eps)}}{\p\al}\Big)^2\Big]
 \, = \, {\frac1{2\leps t}}\Bbb E\Big[-\frac{\p^2l_{12}^{(\eps)}}{\p\al^2}\Big]= d_{22}.
 \eeao
We abbreviate $T_i:=T(X_i,Y_i)$ and find
 from Lemma~4.4 in \cite{EsK2} that
 $\mu_T:=\Bbb E(T_i)=\frac{\al\log2}{\thet^2}-\frac1{\al+\thet}$.
 Then by \eqref{score1} and \eqref{score2} we find
 \beao
m_{13}&=&{\frac1{2\leps t}}\Bbb E\Big[\Big(\frac{\p
l_{12}^{(\eps)}}{\p\log c}\Big)\Big(\frac{\p l^{(\eps)}}{\p\thet}\Big)\Big]\\
& =& {\frac1{2\leps t}}\Bbb E\Big[(n-2\leps t)\Big(-\lepsd t
\frac{\al\log2}{\thet^2} +\frac{n^\|}{\al+\thet}
+\sum_{i=1}^{n^\|}(T_i-\mu_T) + n^{\|} (\frac{\al\log 2}{\theta^2}-\frac1{\alpha+\theta})\Big) \Big]\\
&=& {\frac1{2\leps t}}\Bbb E\Big[(n-2\leps t)\Big(\frac{\al\log 2}{\theta^2}(n^\|-\lepsd t) + \sum_{i=1}^{n^\|}(T_i-\mu_T)\Big)\Big]\\
&=&  \frac{\al\log 2}{{2\leps t}\theta^2} \Bbb E\Big[n (n^\|-\lepsd t)\Big]
+  {\frac1{2\leps t}}\Bbb E\Big[ \Bbb E\Big[ n\sum_{i=1}^{n^\|}(T_i-\mu_T)\Big| n,n^\|\Big]\Big]\\
& = &  \frac{\al\log 2}{{2\leps t}\theta^2} \left(\Bbb E [n n^\|]
- 2\la^{(\eps)}\lepsd t^2\right) \, = \, \frac{\al\log
2}{{2\leps t}\theta^2} \cov(n,n^\|) \, = \,  2
{d}\frac{\al\log 2}{\theta^2},
 \eeao
 where we have used Lemma~\ref{nnd}.
 \beao
 m_{23} &=&{\frac1{2\leps t}} \Bbb E\Big[ \Big(\frac{\p l_{12}^{(\eps)}}{\p\al} \Big)
\Big(\frac{\p l^{(\eps)}}{\p\thet}\Big)\Big]\\
&=& - {\frac1{2\leps t}}\Bbb
E\Big[\Big(\sum_{i=1}^{n}(\log\frac{Z_i}{\eps}-\frac{1}{\al})+\log\eps(n-2\leps
t)\Big)
 \Big( \frac{\al\log2}{\thet^2}(n^\|  -\lepsd t)
+\sum_{i=1}^{n^\|}(T_i-\mu_T)\Big)\Big]\\
&=& -{\frac1{2\leps t}}\Bbb
E\Big[\Big(\sum_{i=1}^{n}(\log\frac{Z_i}{\eps}-\frac{1}{\al})\Big)
\Big(\sum_{i=1}^{n^\|}(T_i-\mu_T)\Big)\Big]\\
&& - \frac{\al\log2}{{2\leps t}\thet^2} \log\eps \, \Bbb
E\Big[(n-2\leps t)(n^\| -\lepsd t)\Big]
\eeao
Now note that the
jumps $(X_i,Y_i)_{i=1,\ldots,n^\|}$ are independent and
independent of all single jumps in either component. Recall that
$T_i=T(X_i,Y_i)=T(X_i/\eps,Y_i/\eps)$, where the last equality is
easily checked. Hence, the right hand side above reduces to
\beao
&=& -{\frac1{2\leps t}} \, \Bbb E\Big[\sum_{i=1}^{n^\|}(\log\frac{X_i}{\eps}+\log\frac{Y_i}{\eps}-\frac{2}{\al})(T_i-\mu_T)\Big]
- \frac{\al\log2}{{{2\leps t}}\thet^2} \log\eps\, \cov(n,n^\|)\\
&=& -{\frac{\lepsd t}{2\leps t}} \, \Bbb E\Big[\Big(\log\frac{X_1}{\eps}+\log\frac{Y_1}{\eps}-\frac{2}{\al}\Big)
\Big(T_i -\mu_T\Big)\Big]
- {\frac{2\lepsd t}{2\leps t}}
\frac{\al\log2}{\thet^2} \log\eps\\
&=& -{d}
\Big(m + \frac{2\al\log2}{\thet^2} \log\eps\Big)\,.
\eeao
Finally, recalling $D=(d_{ij})_{1\le i,j\le 3}$,
 \beao
 m_{33}&=&{\frac1{2\leps t}}\Bbb E\Big[\Big(\frac{\p l^{(\eps)}}{\p
 \thet}\Big)^2\Big]=-{\frac1{2\leps t}} \Bbb E\Big[\frac{\p^2 l^{(\eps)}}{\p
  \thet^2}\Big]= d_{33}.
  \eeao
  \halmos
\end{document}